\documentclass[aps,prx,twocolumn,superscriptaddress,showpacs,floatfix,longbibliography]{revtex4-1}
\usepackage{amsmath,amssymb,graphicx}

\usepackage[utf8]{inputenc}
\usepackage[T1]{fontenc}
\usepackage{xcolor}

\IfFileExists{newtxtext.sty}
   {\usepackage{newtxtext,newtxmath}}
   {\IfFileExists{stix.sty}
      {\usepackage{stix}}
      {\IfFileExists{mathptmx.sty}
      {\usepackage{mathptmx}}{} } }

\usepackage{textcomp}

\usepackage{bm}

\usepackage[normalem]{ulem}
\usepackage[makeroom]{cancel}
\DeclareRobustCommand\redsout{\bgroup\markoverwith{\textcolor{red}{\rule[0.5ex]{2pt}{0.4pt}}}\ULon}
\newcommand\Ccancel[2][black]{
    \let\OldcancelColor\CancelColor
    \renewcommand\CancelColor{\color{#1}}
    \cancel{#2}
    \renewcommand\CancelColor{\OldcancelColor}
}

\IfFileExists{siunitx.sty}{\usepackage{booktabs,siunitx}}{}

\usepackage{color}
\definecolor{LinkColor}{rgb}{0.256,0.439,0.588}
\usepackage{hyperref}

\renewcommand{\vec}[1]{\mathbf{#1}}

\usepackage{pifont}

\providecommand{\bv}[1]{\bm{\mathrm{#1}}}

\providecommand{\w}{\omega}
\providecommand{\wn}{\omega_n}
\providecommand{\W}{\Omega}
\providecommand{\Wm}{\Omega_m}
\providecommand{\q}{\bv{q}}

\providecommand{\vf}{\upsilon_\text{F}}
\providecommand{\vfv}{\bv{\vf}}
\providecommand{\kf}{k_\text{F}}

\providecommand{\kf}{k_\text{F}}
\providecommand{\Tc}{T_\text{c}}
\providecommand{\SigmaT}{\Sigma_\text{T}}
\providecommand{\SigmaQ}{\Sigma_\text{Q}}
\providecommand{\EF}{E_\text{F}}
\providecommand{\wF}{\omega_\text{F}}
\providecommand{\wb}{\omega_\text{b}}

\renewcommand{\q}{\bv{q}}

\providecommand{\gb}{\bar{g}}

\providecommand{\tp}{2\pi}
\providecommand{\tpp}{\left(2\pi\right)}

\providecommand{\Sg}{\Sigma}

\begin{document}

\title{ Identification of non-Fermi liquid fermionic self-energy from quantum Monte Carlo data}

\author{Xiao Yan Xu}
\email{wanderxu@gmail.com}
\affiliation{Department of Physics, University of California at San Diego, La Jolla, California 92093, USA}

\author{Avraham Klein}
\affiliation{School of Physics and Astronomy, University of Minnesota, Minneapolis, MN 55455, USA}

\author{Kai Sun}
\affiliation{Department of Physics, University of Michigan, Ann Arbor, Michigan 48109, USA}

\author{Andrey V. Chubukov}
\affiliation{School of Physics and Astronomy, University of Minnesota, Minneapolis, MN 55455, USA}

\author{Zi Yang Meng}
\email{zymeng@hku.hk}
\affiliation{Department of Physics and HKU-UCAS Joint Institute of Theoretical and Computational Physics, The University of Hong Kong, Pokfulam Road, Hong Kong SAR, China}
\affiliation{Beijing National Laboratory for Condensed Matter Physics and Institute of Physics, Chinese Academy of Sciences, Beijing 100190, China}
\affiliation{Songshan Lake Materials Laboratory, Dongguan, Guangdong 523808, China}

\begin{abstract}
  Quantum Monte Carlo (QMC) simulations of correlated electron systems provide unbiased information about system behavior at a quantum critical point (QCP) and can verify or disprove the existing theories of non-Fermi liquid (NFL) behavior at a QCP.  However, simulations are carried out at a finite temperature, where quantum-critical features are  masked by finite temperature effects. Here we present a theoretical framework within which it is possible to
  separate thermal and quantum effects and extract the information about NFL physics at $T=0$. We demonstrate our method  for  a specific example of 2D fermions near a Ising-ferromagnetic QCP. We show that one can
  extract  from QMC data  the  zero-temperature form of fermionic self-energy $\Sigma (\omega)$ even though the leading contribution to the self-energy comes from thermal effects. We find that  the   frequency dependence of $\Sigma (\omega)$  agrees well with the analytic form obtained within the Eliashberg theory of dynamical quantum criticality, and  obeys $\omega^{2/3}$ scaling at low frequencies. Our results  open up an avenue for QMC studies of quantum-critical metals.
\end{abstract}

\date{\today}

\maketitle
\section{introduction}
\label{sec:intro}
Understanding non-Fermi liquid (NFL) behavior near a metallic quantum-critical point (QCP) remains one of the most ambitious goals of the studies of interacting electrons. 
Examples of systems evincing metallic quantum criticality include
fermions in spatial dimensions $D \leq 3$ at the verge of either spin-density-wave, or charge-density-wave, or nematic order, 2D fermions at a half-filled Landau level, quarks at the verge of an instability to color superconductivity, and several SYK-type models with either electron-electron or electron-phonon  interaction~\cite{sachdevbook,Hertz1976,Moriya1985,Lee1989,Millis1992,Millis1993,Altshuler1994,Polchinski1994,Nayak1994,son,son2,Abanov2000,Oganesyan2001,Abanov2001,Stewart2001,Abanov2003,Metzner2003,Custers2003,Abanov2004,Chubukov2005a,DellAnna2006,Rech2006,Maslov2006,Leohneysen2007, Lee2009,Maslov2009,Metlitski2010a,Metlitski2010b,Metlitski2010,Mross2010,Holder2015,Holder2015a,Wang2016,Wang2017a,Lee2018,HQYuanFMQCP2020,Torroba2019,
  Wu2019,Esterlis2019,Wang2020,GPPan2020,YichenXu2020,Damia2020}. At a  QCP, fluctuations of the corresponding bosonic order parameter become soft. The fermion-fermion interaction, mediated by these soft fluctuations, yields a fermionic self-energy $\Sigma (\omega) \propto |\omega|^{a}$ with $a <1$.
The real and imaginary parts of this self-energy are comparable in magnitude
and both are larger than $\omega$ at 
low frequencies.
This implies that the damping of quasiparticles remains comparable to their energy even infinitesimally close to the Fermi surface, in variance with the central paradigm of 
Landau's theory of a Fermi liquid (FL).  Studies of NFL became the mainstream of research on correlated electrons  after a series of discoveries of high-temperature superconductors, which display unconventional metallic properties in the normal state~\cite{Stewart2001,Leohneysen2007,Hartnoll2016book,Keimer2015}. In most of these materials, superconductivity borders other ordered phases with either spin or charge order.
There are also multiple overlaps between the behavior of 
fermions at a QCP and 
high-energy physics and string theory~\cite{Hartnoll2016book,Maldacena2016}. 

\begin{figure*}[t]
  \centering
  \includegraphics[width=0.99\hsize]{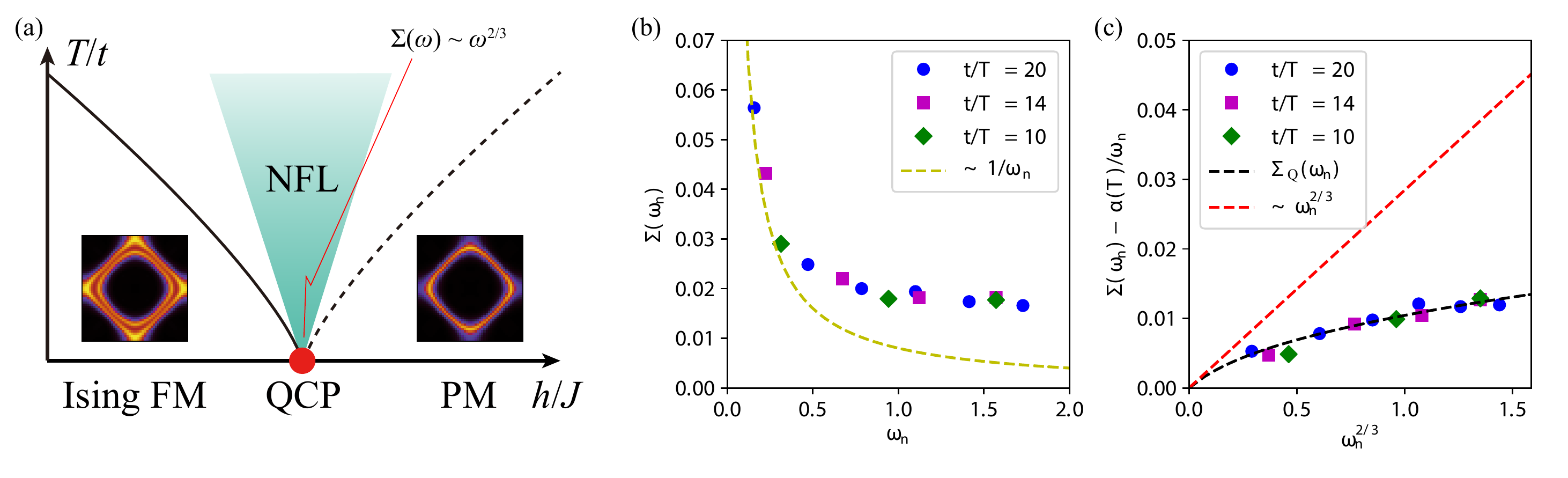}
  \caption{\textbf{Identification of the non Fermi liquid.} (a) Schematic phase diagram of a FM (2+1)D QCP, adapted from Ref.~\cite{xu2017non}. (b) Fermionic self-energy at a FM (2+1)D QCP, calculated by QMC simulation, adapted from Ref.~\cite{xu2017non}. Here we focus on the point on the Fermi surface with Fermi wavevector along the $x$ direction. The QMC self-energy appears to have a leading term of the form $1/\wn$. (c) Quantum part of fermionic self-energy at a FM (2+1)D QCP.
    The black dashed line shows the theoretical prediction of the zero-temperature fermi-self energy, while the red dashed line marks the low-frequency asymptotic form. We
    emphasize that the theory is parameter free, and all
    system parameters, e.g, Fermi velocities, are determined separately from the model or QMC measurement.  }
  \label{fig:fig1}
\end{figure*}

In recent years, several analytical approaches 
have been developed to study NFL behavior at a QCP. These approaches are based 
on effective fermion-boson models, in which 
soft fluctuations of a specific order parameter serve as the source of NFL behavior.
The long-standing goal of these studies is to find the  functional form of $\Sigma (\omega)$ at a QCP and extract the exponent $a <1$ from its small $\omega$ behavior.
One-loop calculations show that $\Sigma (\omega)$ does become singular at a QCP, e.g., in 2D
at a transition to nematic or Ising ferromagnetic order with momentum 
$Q = 0$, it scales at the lowest frequencies as $\omega^{2/3}$ ($a =2/3$).
Whether this behavior extends beyond one-loop is a more tricky issue.  
Power counting arguments indicate that higher-order terms in the loop-expansion for the self-energy reproduce the $\omega^{2/3}$ scaling form~\cite{Altshuler1994}. 
However, detailed calculations reveal that additional $(\log{\omega})^n$ factors appear, and that $n$ increases with the loop order 
\cite{Lee2009,Metlitski2010a,
  Metlitski2010,Holder2015,Holder2015a,Lee2018}.
Such logarithms imply that at low enough frequencies, $\omega \ll \omega_{\text{mod}}$,  $\Sigma(\omega)$ gets modified from its one-loop form.
As a further complication,  the same interaction that gives rise to NFL behavior also gives rise to superconductivity at a non-zero $\Tc$, so normal state self-energy holds only at $\omega > \Tc$.   
It is difficult to extract from analytical studies 
whether $\omega_{\text{mod}}$ is  larger or smaller than $\Tc$, i.e., whether
the modification of $\Sigma (\omega)$  from higher-order processes is relevant for a metal, which displays superconductivity near a QCP,  or only for a putative normal state at $T=0$. This uncertainty
has triggered an interest in
independent numerical studies of the behavior of fermions at a metallic QCP.

Numerical methods for itinerant fermions near a QCP have witnessed great progress in recent years, and at present one can  analyze quantum criticality via reliable large-scale numerical simulations~\cite{Berg2019,Xu2019}. In particular, it has been found that designer models of fermion-boson models offer a pathway to access fermionic QCPs while avoiding the notorious sign-problem in large-scale quantum Monte Carlo (QMC) simulations. Such models have been implemented in several
simulations, studying nematic~\cite{Schattner2016,xu2016topo}, ferromagnetic~\cite{xu2017non}, antiferromagnetic~\cite{ZHLiu2018,ZHLiu2019,Schattner2016a,Gerlach2017,Bauer2020}, gauge field~\cite{XYXu2019,ChuangChen2019,Assaad2016,Gazit2016,Gazit2019,ChuangChen2020} and Yukawa-SYK-type~\cite{GPPan2020} QCPs. The focusing  on a particular soft boson
offers an unbiased numerical test for either a $Q=0$ or a finite $Q$ analytical theory of metallic quantum criticality.
The mutual inspiration and dialogue between numerical and theoretical communities, arising from these studies, has also stimulated progress along the numerical front (SLMC~\cite{XYXu2017} and EMUS~\cite{ZHLiu2019EMUS} are successful examples of this).

Sign-problem free QMC has its own limitations as well.
To avoid superconductivity and finite size effects, 
simulations are done at a
finite $T$ which is not the smallest energy scale in the system, such that
on a Matsubara axis the
fermionic self-energy $\Sigma (\wn)$ is a function of a discrete Matsubara frequency 
$\wn = (2n+1)\pi T$
. (The self energy also has a momentum dependence $\Sigma = \Sigma(\wn, \mathbf{k})$, but here and henceforth we suppress this notation for clarity, except where needed.)
At nonzero $T$ it can generally be expressed as $\Sigma (\wn) = \SigmaT (\wn) + \SigmaQ (\wn)$, where the ``thermal'' part
$\SigmaT (\wn)$ is the  contribution from static thermal fluctuations
and the ``quantum'' part $\SigmaQ (\wn)$ is the contribution from dynamical bosonic fluctuations.  At $T=0$,  $\wn$ is a continuous variable, $\SigmaT = 0$, and $\Sigma = \SigmaQ (\wn)$ is a NFL self-energy at a QCP. 
However, at a finite $T$, 
the self-energy differs from its $T=0$ form,  and the presence of $\SigmaT (\wn)$ can mask the behavior associated with $\SigmaQ (\wn)$. Besides, at a finite $T$, $\SigmaQ (\wn)$ 
also generally differs from its $T=0$ form. We note that in the Yukawa-SYK model these finite temperature effects have recently been analyzed using an emergent conformal (reparametrization) symmetry of the low-energy theory, which automatically incorporates thermal and quantum effects~\cite{GPPan2020}. However, the treatment of the present critial FS model without confmal symmetry requires separate analyses of $\SigmaQ$ and $\SigmaT$.

The main purpose of this paper is to provide the  method to disentangle $\SigmaT$  and $\SigmaQ$ from  QMC data for the self-energy. Our approach is based on three observations. 

\begin{itemize}
\item First, to study QC behavior one should avoid
the effect of
fluctuations from fermions with energies of order of the bandwidth, such
as would lead to e.g. Mott physics.
For this, the effective femion-boson coupling (labeled $\bar g$ in the text) should be
much smaller than the bandwidth $W$.
In systems with a large Fermi surface, $W$ is comparable to the Fermi energy, so the necessary condition is $\gb \ll \EF$.

\item Second, at small ${\bar g}$, there is a wide range of frequencies $\wn \ll \EF$, for which  
$\wn$ is much larger than $\Sigma (\wn)$.
In this range,  the thermal self-energy has a simple form,
valid for finite temperatures and frequencies $\wn \gg \Sg$,
$\SigmaT(\wn) \approx \alpha (T)/\wn$ up to logarithmic corrections, i.e., 
$\wn \SigmaT (\wn) = \alpha (T)$ is approximately independent of $\wn$.
\item Third, in the same range,  
  $\SigmaQ (\wn)$ still has NFL form  and is well approximated by the one-loop,  $T=0$ expression, modulo that $\wn$ is discrete.
\end{itemize}

By considering the above points, one arrives at the following conclusion:
if a QMC study is performed at ${\bar g} \ll \EF$ and provides  data for $\Sigma (\wn)$ for a  substantial number of Matsubara points in the range $\wn \gg \Sigma (\wn)$,
it is possible to extract $\SigmaQ$ from the data by the following simple procedure. First, extract 
(the approximatly constant) $\alpha(T)$ from the data by fitting $\wn \Sigma (\wn)$ by a continuous function of frequency and extrapolating to zero frequency, where it is equal to $\alpha(T)$
because $\wn \SigmaQ (\wn)$ extrapolates to zero. 
Once $\alpha_T$ is known,
subtract $\SigmaT(\wn) = \alpha (T)/\wn$ from the full $\Sigma (\wn)$
and obtain $\SigmaQ (\wn)$, which, as we said,  should have the same form as $T=0$ 
self-energy.
For a more accurate separation of $\SigmaQ$ from $\SigmaT$ include the slow frequency dependence of $\alpha(T)$ in the fitting procedure, which is still quite straightforward to do, as we will show later.

\begin{figure}[t]
  \centering
	\includegraphics[width=0.95\hsize]{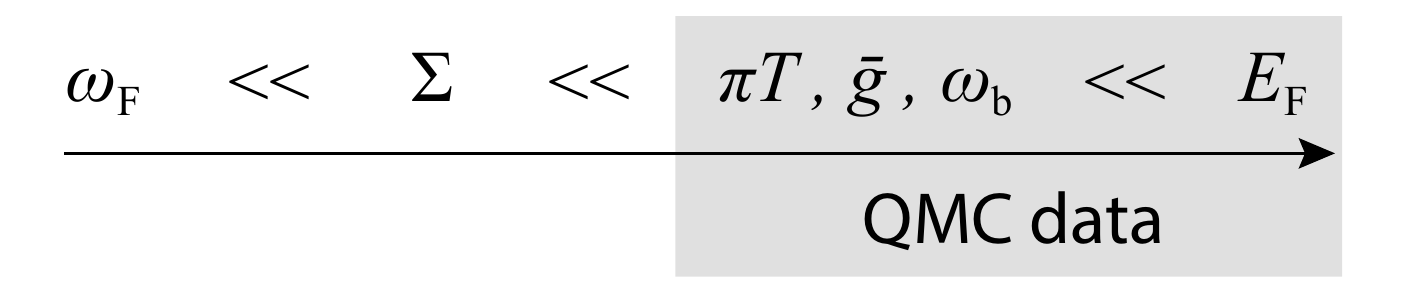}
	\caption{\textbf{Schematic representation of the energy scale relevant in our QMC study.}}
	\label{fig:energyscale}
\end{figure}

We apply our strategy to a metal near an Ising ferromagnetic QCP. We show the schematic phase diagram in  Fig.~\ref{fig:fig1}(a). It contains regions of a paramagnetic metal (PM) and  an ordered
Ising-ferromagnet (Ising FM), separated by a QCP.  Right above the QCP, there is a  region of small $T$, where the system displays truly NFL behavior, i.e. $\Sigma (\wn)$ is non-analytic and larger than $\wn$.  At higher $T$,
$\Sigma (\wn)$ becomes smaller than $\wn$,  yet the self-energy still has non-FL form, and, by our reasoning, its quantum part, $\SigmaQ (\wn)$ should be almost the same as at $T=0$.  In Fig.~\ref{fig:fig1}(b) we show the full self-energy, obtained in QMC simulation, and in  Fig.~\ref{fig:fig1}(c) we show $\SigmaQ (\wn)$, extracted using the
approximate
procedure outlined above.  The black line in Fig.~\ref{fig:fig1}(c) is the analytical one-loop  result for the self-energy at a QCP at $T=0$ . We see that the data for all $\wn$
nicely fall onto this curve.  At small $\wn$, the analytic one-loop self-energy behaves as $\omega^{2/3}_n$, and the fact that QMC data fall onto
the $T=0$ curve  implies that
the QMC data are consistent with $\wn^{2/3}$  scaling at the lowest $\wn$ at a  QCP.  The deviation from $\wn^{2/3}$ scaling in the analytical formula (Eq. \eqref{eq:Sg-Q-MET-Scaling} in the text) is
due to two reasons. First, for the model used for QMC simulations, the bosonic propagator $D(q, \Omega)$ contains a regular $\Omega^2$ term along with the Landau damping term, $\Omega/q$. 
When this term becomes relevant, 
$\SigmaQ (\wn)$  tends to saturate. Second, even when the Landau damping term dominates, 
the $\omega^{2/3}$ form 
is the low frequency limit of a more complicated 
function $\SigmaQ(\wn) \propto \wn^{2/3}
\mathcal{U}\left(\wn/\omega_b\right)$, and $\omega^{2/3}$ behavior holds only when $\omega_n \ll \wb$, i.e., $\mathcal{U}(z) \approx \mathcal{U}(0)$.
The crossover frequency $\wb \sim ({\bar g} \EF)^{1/2}$ (see Eq. \eqref{eq:wb-def} below).  In our simulations this $\wb$ is much larger than the upper boundary of NFL behavior, $\wF \sim {\bar g}^2/\EF$, but
is still much smaller than $\EF$. Accordingly, most of our $\wn$ fall into $\wn > \wb$, 
where $\SigmaQ(\wn)$ differs from $\wn^{2/3}$.
We emphasize that $\SigmaQ(\wn)$ has a NFL form 
regardless of the ratio $\wn/\wb$. 
  Fig. \ref{fig:energyscale} presents a summary of
  the relevant
  energy scales in our QMC study.

It is instructive to compare our results with recent analysis of QMC data for similar models.
Ref. \cite{KleinUn2020} demonstrated that a rather flat dispersion of $\Sigma (\wn)$, obtained in QMC simulations, is reasonably well reproduced by $\Sigma (\wn) = \SigmaT(\wn) + \SigmaQ (\wn)$, where both are computed analytically within a metallic QC theory.  For that study, a larger coupling ${\bar g} \sim \EF$ was used to increase the magnitude of the self-energy. The discrepancy between the analytic and QMC self-energies in Ref. \cite{KleinUn2020} was  about  20\%.  This was small enough to see  that analytic and QMC self-energies have similar dispersion, but still too high to reliably extract $\SigmaQ (\wn)$ from the QMC data.  For the current study, ${\bar g}$ is smaller, and typical $\Sigma(\wn)/\wn$ is roughly 5 times smaller than in that work. In this situation, we argue that the QC form of $\SigmaQ (\wn)$ can be extracted from the data.

The structure of the paper is the following.  In Sec.~\ref{sec:SecII} we describe the lattice model for which the QMC simulations have been performed,
and present the numerical results for the self-energy.  In Sec.~\ref{sec:SecIII} we present the analytical results  for the self-energy within the self-consistent one-loop analysis.
In Sec.~\ref{sec:SecIV} we extract $\SigmaQ (\wn)$ from QMC data and show that for all $n >0$ it falls onto  the analytic, $T=0$ form of $\SigmaQ (\wn)$.
In Sec.~\ref{sec:SecV} we summarize the results and discuss the implication of this work to other QC cases studied in QMC simulations.
We argue that the computational scheme that we proposed can be used as a generic method to extract NFL self-energy at a QCP and can be further extended to study more subtle effects, e.g., the flow of the dynamical exponent $z$.

\section{Results}
\begin{figure}[t]
  \centering
	\includegraphics[width=0.95\hsize]{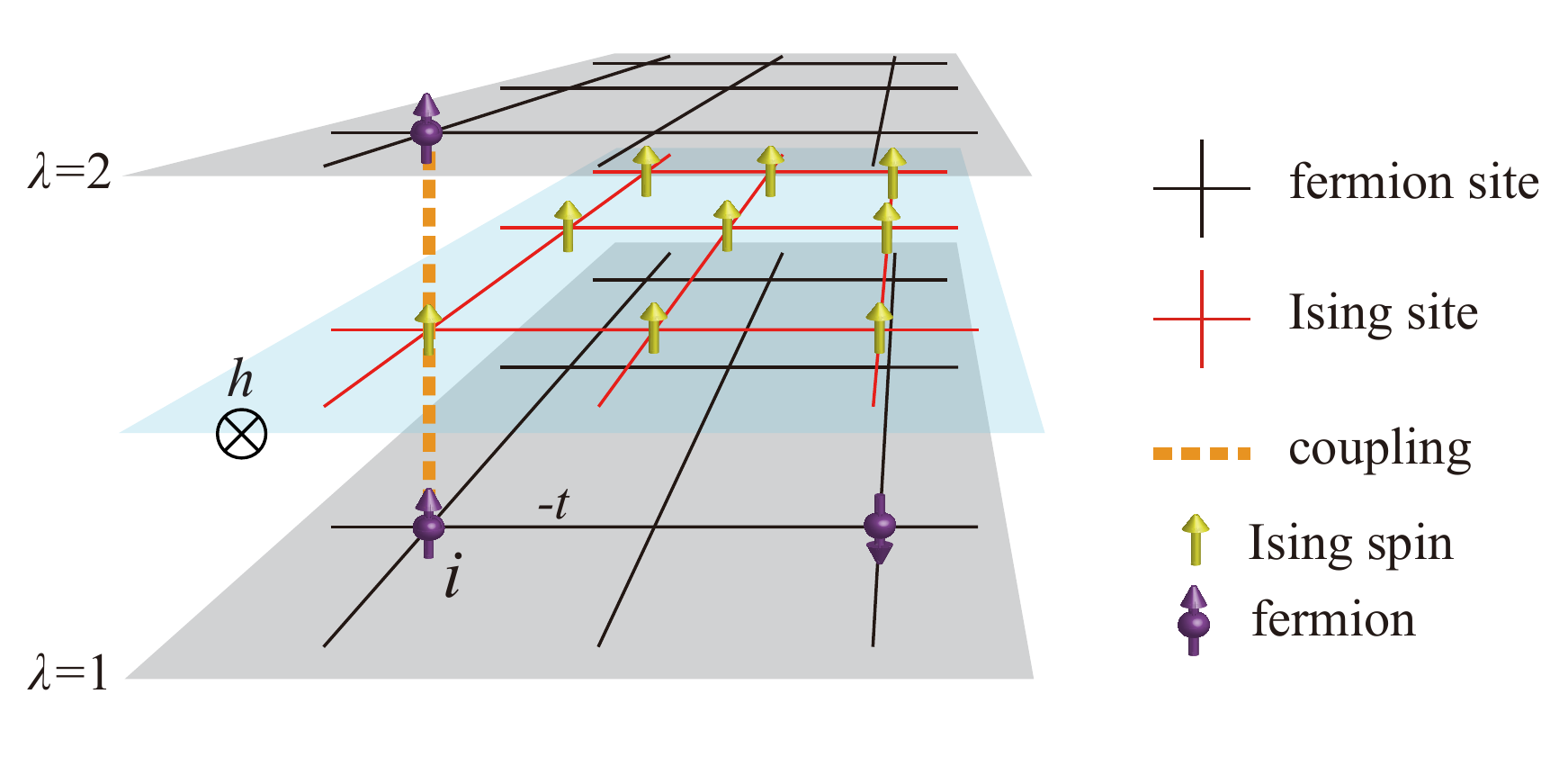}
	\caption{\textbf{Schematic representation of the Ising FM model used in QMC simulations.} Adapted from Ref.\cite{xu2017non}.}
	\label{fig:fig2}
\end{figure}

\subsection{The lattice model, phase diagram and QMC self-energy}
\label{sec:SecII}
As shown in Fig.~\ref{fig:fig2}, we consider a model describing Ising FM fluctuations coupled to a Fermi surface~\cite{xu2017non}. The model is implemented on a square lattice with Hamiltonian $\hat{H}=\hat{H}_\text{f}+\hat{H}_\text{s}+\hat{H}_\text{sf}$ and each part reads
\begin{eqnarray}
  \hat{H}_{\text{f}}&=&-t\sum_{\langle ij\rangle\lambda\sigma}\left(\hat{c}_{i\lambda\sigma}^{\dagger}\hat{c}_{j\lambda\sigma}+h.c.\right)-\mu\sum_{i\lambda\sigma}\hat{n}_{i\lambda\sigma} \nonumber\\
  \hat{H}_{\text{s}}&=&-J\sum_{\langle ij\rangle}\hat{s}_{i}^{z}\hat{s}_{j}^{z}-h\sum_{i}\hat{s}_{i}^{x} \nonumber\\
  \hat{H}_{\text{sf}}&=&-\xi\sum_{i}s_{i}^{z}(\hat{\sigma}_{i1}^{z}+\hat{\sigma}_{i2}^{z})
\label{eq:eq2}
\end{eqnarray}
where $\hat{H}_\text{f}$ describes two layers (or orbitals, $\lambda$=1,2) of spinful ($\sigma=\uparrow,\ \downarrow$) fermions with nearest-neighbor hopping on a square lattice, and the chemical potential $\mu$ tunes the size of bare Fermi surface.
The bare fermion dispersion dictated by $\hat{H}_\text{f}$ is $\epsilon(\vec{k})=-2t(\cos(k_x)+\cos(k_y))-\mu$
and the bandwidth is $W=8t$.  $\hat{H}_\text{s}$ represents a transverse field Ising model on the same lattice, where by tuning $T$ and $h/J$ an Ising ferromagnet (FM) to paramagnet (PM) transition can be obtained.
The onsite coupling term $\hat{H}_\text{sf}$ between the fermions and Ising spins, mediates a fermion-fermion interaction, establishing a metallic system with ferromagnetic fluctuations. We presented the schematic phase diagram in Fig.~\ref{fig:fig1}(a). In the analysis which follows, we focus on the model parameters $\{t=1,\  \mu=-0.5t,\ J=1,\ \xi/t=1\}$, for which we find a FM QCP at $h_\text{c}/J \approx3.270(6)$. The parameters associated with the fermiology for these parameters are listed in Table~\ref{tab:ekpara}.

As shown in Ref.~\cite{xu2017non}, our model gives rise to a FM-QCP.
However, the bare numerical fermionic self-energy data from QCP, as shown in Fig.~\ref{fig:fig1}(b) shows a behavior distinctively different from the expected NFL  $\Sigma(\wn) \propto \wn^{2/3}$.
At low frequency, the self-energy shows an unusual upturn instead of going to zero. Such a upturn in the imaginary part of fermionic self-energy, in the usual numeric setting, implies a gap opening on the Fermi surface. However, our data of the fermionic Green's function
does not show a well-formed gap on the  FS.
Similar behavior of the numerical NFL self-energies have also been observed in other cases including nematic- and AFM-QCPs~\cite{Schattner2016,ZHLiu2019}.
As discussed in the introduction, the rest of this paper is devoted to an analysis of the self-energy data in Fig. \ref{fig:fig1}(b), and to understanding how to disentangle the thermal and quantum parts of the self energy, as shown in Fig. \ref{fig:fig1}(c).
\begin{table}
\caption{Parameters of the fermiology. Here $\mathcal{V}_\text{F} = k_\text{F}/\upsilon_\text{F}$ denotes the density of states.}
\begin{tabular}{|c|c|c|c|c|c|}
  \hline
  & $(k_{x},k_{y})$ & $k_{\text{F}}$ & $\upsilon_\text{F}$ & $\mathcal{V}_\text{F}$ & $E_\text{F}$\tabularnewline
  \hline
  \hline
  $\theta=0$ & (2.42,0) & 2.42 & 1.32 & 1.83 & 1.60\tabularnewline
  \hline
  $\theta=\frac{\pi}{4}$ & (1.44,1.44) & 2.04 & 2.81 & 0.73 & 2.87\tabularnewline
  \hline
\end{tabular}
\label{tab:ekpara}
\end{table}

\subsection{Analytic self-energy at Ising-FM QCP}
\label{sec:SecIII}

We begin with a brief review of the diagrammatic theory for interacting fermions near the ferromagnetic QCP. As the derivations of the electron-boson models and their relationship to itinerant QCP and NFL physics as well as superconductivity are scattered over numerous research papers and reviews encompassing decades of work, assiduous readers are suggested to directly consult these references~\cite{Hertz1976,Moriya1985,Millis1992,Millis1993,Altshuler1994,Bonesteel1996,Abanov2000,Abanov2004,DellAnna2006,Rech2006,Abanov2001,Abanov2003,Metzner2003,Leohneysen2007,Lee2009,Metlitski2010a,Metlitski2010,Metlitski2010b,Lederer2017,Lee2018,Maslov2010,Fradkin2010,Wang2016,Raghu2015,Metlitski2015,Lederer2015,Wang2017a,Damia2020}. Here we will keep the our derivation concise and try to be self-contained.

\begin{figure}
  \centering
  \includegraphics[width=0.6\hsize]{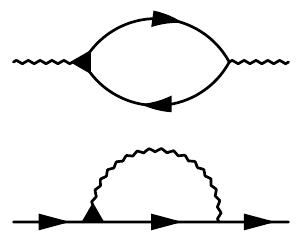}
  \caption{\textbf{The diagrammatic representation of bosonic self energy $\Pi(q)$ and fermionic self energy $\Sg(k)$.}}
  \label{fig:fig3}
\end{figure}

To understand the situation described in Eq.~\eqref{eq:eq2} of itinerant electrons coupled to critical bosonic fluctuations, we can encode the dynamics of bosons and fermions in their propagators,
\begin{equation}
  \label{eq:G-def}
  G(k) = \left(\text{i}\w_n + \text{i}\Sigma(k) - \epsilon(\mathbf{k})\right)^{-1},
\end{equation}
and
\begin{equation}
  \label{eq:D-def}
  D(q) = D_0\left(M^2_0 + |\mathbf{q}|^2 + c^{-2}\W_m^2 + \Pi(q)\right)^{-1},
\end{equation}
where $k = (\w_n,\mathbf{k}),q=(\W_m,\mathbf{q})$ are three-vectors with $\wn=(2n+1)\pi T$ and $\Wm = 2m\pi T$ the fermionic and bosonic Matsubara frequencies respectively, $\epsilon(\mathbf{k})$ is the dispersion from Sec. \ref{sec:SecII}, $M_0^2$ represents the bare distance to the QCP before the interaction is turned on (in the QMC it is controlled by the transverse magnetic field, $M_0 = M_0(h)$), and $\Sg,\Pi$ are respectively the fermionic and bosonic self-energies. Both self-energies are represented by a diagrammatic series in $\gb = (\frac{\xi}{2})^2 D_0$. The series is depicted pictorially in Fig.~\ref{fig:fig3}, where solid and wiggly lines are the full propagators $G(k),D(q)$ and the triangles are fully dressed vertices. In general, it is not justified to neglect the vertex corrections. However, it is customary to split the corrections into two types: those coming from fermions away from the Fermi surface (``high energy'' fermions on the scale of the bandwidth $W$), and those coming from near the Fermi surface. The high energy contributions just give some static corrections to an effective low-energy theory, which can be absorbed into an effective renormalized coupling $\gb$. The condition for the smallness of these corrections is weak coupling,
\begin{equation}
  \label{eq:eq5}
  \gb \ll \EF.
\end{equation}
This condition is valid away from the QCP, i.e. when $M_0^2 \sim \kf^2$. In the low energy theory, at low enough temperatures and frequencies it is not justified to neglect vertex corrections. However, those vertex corrections that contribute to $\Sg(k)$,
can be neglected if we are in a regime where $|\Sg(\w_n)| \ll \w_n$.
As shown in Fig.~\ref{fig:fig1} (b), the lowest fermionic frequency in our QMC simulation is $\omega_0=\pi T=0.157$ with $T=t/20$ and the corresponding fermionic self-energy $|\Sigma(\kf,\omega_0)|=0.058$, so this condition is
satisfied. A longer discussion on this is presented in another work by some of us~\cite{KleinUn2020}.

In our QMC study we are always
  in the
  regime
  $|\Sigma(\wn)| \ll \wn$, and Eq. (\ref{eq:eq5}) is obeyed, so without further discussion we will assume that vertex corrections are negligible. Then $\Pi,\Sg$ are described by the coupled self-consistent equations,
\begin{align}
  -\text{i}\Sigma(k) &= \gb T \sum_n \int \frac{d^2p}{\tpp^2}G\left(p+k\right)D\left(p\right),\label{eq:1loop-sig} \\
  \Pi(q) &= 2 N_\text{f} \gb T \sum_{n}\int \frac{d^2p}{\tpp^2}G\left(p+q\right)G\left(p\right). \label{eq:1loop-Pi}
\end{align}
Here $N_\text{f}$ is the number of fermion flavors ($N_\text{f} = 2$ in the model of Sec. \ref{sec:SecII}) and the factor 2 in $\Pi$ comes from spin summation.

In principle, Eqs. (\ref{eq:1loop-sig}) and (\ref{eq:1loop-Pi}) have momentum integrals over the entire Brillouin zone, which means that they still include contributions to the self-energies that come from high energies. One of these is a static contribution to $\Pi$. This contribution just renormalizes the mass towards the QCP, i.e. $M_0^2$ in Eq. (\ref{eq:D-def}) is replaced by
\begin{equation}
  \label{eq:mass-deff}
  M^2 = M_0^2 - \Pi(\W_m = 0, \q = 0).
\end{equation}
Thus, $M^2$ can be tuned to a QCP by varying $\gb$, or alternatively by varying $M_0^2$ (this is what is done in the QMC simulations). An additional static contribution renormalizes $D_0$ and we absorb it into $\gb$. There are also static contributions to $\Sg$, but they do not change the critical dynamics so we absorb them into the fermionic dispersion. Then there are dynamical contributions which we will now compute.

Beyond neglecting vertex corrections, we further assume that the fermionic dispersion can be linearized near the FS, which means that the theory describes a low-energy effective theory near the FS. Then, integrating over linearized fermionic dispersion we obtain,
\begin{widetext}
\begin{align}
  \label{eq:Pi-low}
  \Pi(q,\Wm) = 2 \text{i} N_\text{f}\gb T\sum_n\int \frac{d\theta}{2\pi} \mathcal{V}_\text{F}(\theta) \frac{\Theta(\wn+\Wm)-\Theta(\wn)}{\text{i}\W_m-\vf(\theta)q\cos(\theta-\theta_q)},
\end{align}
where $\Theta(x)$ is the step function, the density of states $\mathcal{V}_\text{F}(\theta) = \kf(\theta)/\vf(\theta)$ and $\kf,\vf$ are the Fermi vector and velocity at an angle $\theta$ on  the FS, as given in Tab.~\ref{tab:ekpara}. In Eq. (\ref{eq:Pi-low}), as $|\Sigma(\wn)| \ll \wn$, we neglected contributions from self-energy and assumed that the $\kf$ and $\vf$ vector are approximately parallel. For the fermionic self energy we get,
\begin{flalign}
  \label{eq:sig-low}
  \Sigma(\kf,\wn) \approx \gb T
  \sum_l \int \frac{p dp}{\tp}\frac{\sigma(\w_l)}{\sqrt{\w_l^2+\vf^2(\theta_k)p^2}}
  \frac{1}{M^2+p^2+c^{-2}(\w_n-\w_l)^2+ \Pi(p\hat n(\theta_{k}),\w_n-\w_l)},
\end{flalign}
\end{widetext}
where $\sigma(x)$ is the sign function and
$\hat n(\theta_{k}) = \left.\frac{(-{\vfv}_y,{\vfv}_x)}{\vf}\right|_{\theta=\theta_{k}}$ is an unit vector pointing parallel to the FS at the angle $\theta_{k}$. In a $C_4$ symmetric system we can replace $\hat n$ by $\vfv/\vf$, since the unit vector only determines the value of $\vf(\theta)$ in Eq. \eqref{eq:Pi-low}.

We first evaluate the bosonic self-energy which to leading order is,
\begin{equation}
  \label{eq:Pi-ET}
  \Pi (\Wm, \q) \approx {\bar g} \frac{N_\text{f} \mathcal{V}_\text{F}(\theta_q)}{\pi}\frac{|\W_m|}{\vf(\theta_q)q},
\end{equation}
where the $C_4$ symmetry of the lattice is used to replace $\vf(\theta\pm\pi/2) = \vf(\theta)$ and similarly for $\mathcal{V}_\text{F}$. Next we turn to the fermionic self energy. Plugging Eq. (\ref{eq:Pi-ET}) into Eq. (\ref{eq:sig-low}) yields,
\begin{widetext}
\begin{flalign}
  \label{eq:sig-low-1}
  \Sigma(\kf,\wn) \approx \frac{\gb T}{2\pi}
  \sum_{l}  \int_0^\infty\frac{\sigma(\w_l)}{\sqrt{\w_l^2+\w^2}}
  \frac{ \w^2 d\w}{\w^3 + (\vf^2M^2 + (\vf/c)^{2}(\w_n-\w_l)^2)\w+\w_b^2|\w_n-\w_l|}
\end{flalign}
\end{widetext}
In Eq.~\eqref{eq:sig-low-1} we rescaled momentum to frequency $\w = \vf p$, and hid the explicit angular dependence $\vf = \vf(\theta_k)$ for conciseness. The frequency scale introduced by $\Pi$ is
\begin{equation}
  \label{eq:wb-def}
  \w_\text{b} = \left(\frac{\gb N_\text{f} \kf \vf}{\pi}\right)^{1/2}.
\end{equation}
From Eq. (\ref{eq:sig-low-1}) we can read the relevant frequency scales for $\Sg(\w_n)$. The typical scale of the $\w_l$ sum is $\w_l \sim \w_n$ due to the sign function, i.e. typical internal frequencies are constrained to be on order of the external frequency.

We now show that at finite temperature, but as long as
$|\Sigma(\wn)| \ll \wn$ the fermionic self-energy in Eq.~\eqref{eq:1loop-sig} splits into two parts: thermal and quantum
(for detailed derivations and discussions see e.g. \cite{Abanov2003,DellAnna2006,Wang2017a,KleinUn2020,Damia2020}.) The quantum part recovers the zero-temperature fermionic self-energy, while the thermal part takes on
  a very simple
  form
  and scales as $1/\wn$. Thus, after simply deducting this $1/\wn$ term, the finite-temperature self-energy directly provides the zero temperature behavior of fermions, although the measurement is done at finite temperature, at which thermal fluctuations has a significant contribution. This is one of the key conclusions of this work.
We separate the summation
in Eq. \eqref{eq:sig-low-1} into two parts
\begin{equation}
  \label{eq:SE-MET}
  \Sg(\w_n) = \SigmaT(\w_n,T\neq0) + \SigmaQ(\w_n,T),
\end{equation}
where $\SigmaT$ is the $\omega_l=\wn$ piece of the sum in Eq.~\eqref{eq:sig-low-1}, namely
\begin{align}
  \label{eq:gamma-eq}
  \SigmaT(\w_n) &\approx \frac{
  \gb T }{\tp \w_n}\mathcal{S}\left(\frac{\vf M}{|\w_n|}\right),
\end{align}
where
\begin{equation}
  \label{eq:scaleF-1}
  \mathcal{S}(x) = \frac{\cosh^{-1}(1/x)}{\sqrt{1-x^2}}\approx\left\{
  \begin{array}{ll}
    \log(2/x) & x \ll 1 \\
    \pi/(2x)  & x \gg 1
  \end{array}\right. .
\end{equation}
As $\mathcal{S}(x)$
vanishes rapidly at large $x$, it predicts that $\SigmaT$ only contributes significantly at finite temperature and close enough to the QCP
($\pi T \gtrsim \vf M $). In that regime, as noted in the introduction, $\alpha(T,\wn) = \w_n\SigmaT(\wn)$ depends at most logarithmically on frequency at the smallest $\wn$, $\alpha(T,\wn)\approx \alpha(T)$.

The quantum part includes all other terms in the Matsubara sum. This sum can be approximately replaced by an integral, which immediately recovers
the $T=0$ form of the fermionic self-energy, i.e.,
\begin{equation}
  \label{eq:Sg-Q-MET-Scaling}
  \SigmaQ(\wn) \approx \gb \sigma(\wn)\left(\frac{\wn}{\w_b}\right)^{2/3}\mathcal{U}\left(\frac{\wn}{\w_b} \right),
\end{equation}
with
\begin{widetext}
\begin{equation}
  \label{eq:sig-ET-2}
  \mathcal{U}(z) = \int_0^\infty \frac{dx dy}{4\pi^2} \frac{ y}{y^3 +  (\vf/c)^{2}x^2 y z^{4/3}+x}\left[ \frac{\sigma(x+1)}{\sqrt{1+\left(\frac{x+1}{y}\right)^2z^{4/3}}}-\frac{\sigma(x-1)}{\sqrt{1+ \left(\frac{x-1}{y}\right)^2z^{4/3}}}\right].
\end{equation}
\end{widetext}
The scaling function $\mathcal{U}(z)$ has the following asymptotics,
\begin{equation}
  \label{eq:sig-ET-3}
  \mathcal{U}(z) = \left\{
    \begin{array}{ll}
      \frac{1}{2\pi\sqrt{3}}& z \ll 1 \\
      \frac{1}{24 z^{2/3}} & 1 \ll z \ll z_c \\
      \frac{u_0}{z^{2/3}} & z_c \ll z
    \end{array}\right. .
\end{equation}
where $z_c^{-1} = (\vf/c)^{3/2}$ and $u_0$ is a constant which depends on $\vf/c$. For $\vf/c\ll 1$, $u_0 \approx 1/8$; while for parameters of Sec.~\ref{sec:SecII} ($\vf/c\approx 0.42$), $u_0 \approx 0.1$. Note in the case of our QMC study $z_c \approx 3.7$, so that the
  intermediate regime cannot really be seen.
Eq. \eqref{eq:sig-ET-2} is exactly the formula we used to generate the black line in the Fig.~\ref{fig:fig1}(c), and is the quantum NFL self-energy $\SigmaQ$ of a FM-QCP. It saturates in the large frequency region as shown in the figure, as predicted by Eq.~\eqref{eq:sig-ET-3} in the $z_c \ll z$ limit. Combining Eqs. \eqref{eq:gamma-eq} and \eqref{eq:Sg-Q-MET-Scaling}, we indeed see that the self energy has a thermal $1/\w_n$ term plus the zero temperature quantum self energy.

Let us briefly elaborate on the physics behind the scaling function $\mathcal{U}(z)$.
In Eq. \eqref{eq:sig-ET-2}, the part in the square brackets correspond to the fermionic propagator and the other part in the integral corresponds to the bosonic propagator. Consider the limit $\omega \ll \omega_b$ corresponding to $z\ll 1$.
Expanding for $z \ll 1$ we find that to leading order the terms in the square brackets are a constant, and the $dx$ integral is limited to $0 < x < 1$. Physically this is the statement that the momentum integration ($\int dy$) is only on bosonic momentum parallel to the FS. In addition the $(\vf/c)^2x^2y$ term in the boson propagator is also negligible, which corresponds to the fact that the bare $\W^2$ part of the boson dynamics is irrelevant at low frequency.
Evaluating Eq. \eqref{eq:Sg-Q-MET-Scaling} for $\wn \ll \omega_\text{b}$ we find,
\begin{equation}
  \label{eq:sig-ET-scaling}
  \SigmaQ(\w_n) = \wF^{1/3}|\w_n|^{2/3}\sigma(\w_n) +  \cdots,
\end{equation}
where
\begin{equation}
  \label{eq:wf-def}
  \wF = \frac{\gb^2}{8\pi^23^{3/2}\mathcal{V}_\text{F}\vf^2N_\text{f}},
\end{equation}
Eq.~\eqref{eq:sig-ET-scaling} is the formula used to generate the red dashed line in Fig.~\ref{fig:fig1}(c), as an asymptotic line of the quantum part of the self-energy predicted by Eq.~\eqref{eq:sig-ET-2}. The analysis of $\Sigma$ leading to Eqs. \eqref{eq:sig-ET-scaling} and \eqref{eq:wf-def}, as well as analogous analysis for superconducting self energy, is conventionally termed ``Eliashberg theory'' (ET), due to its similarity to Eliashberg's theory of superconductivity from electron-phonon interactions \cite{Abrikosov1975}.

Now consider the opposite limit, $\omega \gg \omega_\text{b}$, corresponding to $z \gg 1$. For simplicity let's assume $\vf/c\ll 1$. In that case the term in the square brackets, corresponding to the fermionic propagator, is not constant, and the bulk of the contribution to $u_0$ is given by the range $1 < x < \infty$. Physically this means that scattering is not confined to be parallel to the FS and is two dimensional, although it is still confined to be near the FS. It is instructive to compute the subleading term for small $z$. After some algebra, one finds that this contribution is also given by 2D scattering, and gives $(2\pi/\sqrt{3})\mathcal{U}(z) \approx 1 - 0.73 z^{1/3}$. This means that for $z\sim 1$, $\Sigma_0$ is reduced by a factor of almost 4 from the expected value if one considers only the leading contribution. This is the reason that the deviation from the asymptotic red line in Fig. \ref{fig:fig1} (c) is so large. At even larger $z$, the $\Omega^2$ term in the bosonic propagator begins to contribute, which just modifies the high-frequency behavior of the self-energy. However, the deviations from $\omega^{2/3}$ scaling occur already at $z \sim 1$.
We term the theory which accounts for both high-frequency modifications and the finite temperature corrections of Eq. \eqref{eq:gamma-eq} a modified Eliashberg theory (MET).

\subsection{Analysis of QMC data}
\label{sec:SecIV}

Now we turn to the QMC data analysis. We study the fermionic self-energy from the FM-QCP model described in Sec.~\ref{sec:SecII} and compare the QMC data with the MET in Sec.~\ref{sec:SecIII}.

Let us begin by going through the relevant physical parameters in the QMC data. We normalize all quantities by the hopping energy $t=1$, see Sec. \ref{sec:SecII} for details. By tuning the transverse field $h$, Ref.~\cite{xu2017non} was able to extract QMC data for different $T$ above the QCP, and also deep in the disordered phase, where the self-energy should have a FL form $\Sg(\w_n) \propto \w_n$ at low frequencies. We concentrate on the data at the QCP.
The parameters from Sec. \ref{sec:SecII} imply a bare $\gb = 1/4$ which is much smaller than
$\EF \approx 1.6$ implying the QMC is in the weak coupling regime (we remind that all energies are quoted in units of the hopping). The bosonic propagator in QMC was found to agree well~\footnote{As shown in Ref.~\cite{xu2017non}, it turns out the bosonic propagator is dominated by the bosonic self-energy part, with a small finite anomalous dimension in $q^2$ and $\Omega^2$ terms, it will not change the main results of this paper.} with Eqs. (\ref{eq:D-def}), (\ref{eq:mass-deff}) and (\ref{eq:Pi-ET}), i.e.
\begin{equation}
  \label{eq:Dqmc-def}
  D(q) = D_0(M^2(T) + q^2 + c^{-2}\W_m^2 + \Pi(q))^{-1},
\end{equation}
with $D_0 = 1, M^2(T) = 0.13 T^{1.48}, c = 3.16$, i.e. the measured $D_0$ agrees with the bare one, all obtained from the bosonic propagator data in Ref.~\cite{xu2017non}. In addition, it was found that $\Sg(\w_n) \ll \w_n$ for all temperatures and Matsubara frequencies that were obtained. Thus, we may expect that corrections to the bare $\gb$ are small, and the renormalized $\gb$ which is an input to MET is at the order of the bare one. Under this condition, the relevant scales for $\Sg$ are
\begin{equation}
  \label{eq:scales}
  \w_\text{b} = 0.71,\  
  \wF = 2.38\times 10^{-5}.
\end{equation}
The temperatures we analyze are $T=0.05\ldots0.1$, which implies the first Matsubara frequency is $\pi T = 0.16\ldots0.31$, see the schematics of energy scale in Fig.~\ref{fig:energyscale}. Thus, $\wF$ is completely irrelevant as is verified by the fact that the self-energy is always small.
As we discussed in Sec.~\ref{sec:intro}, the QMC self-energy appears to have a leading term of the form
\begin{equation}
  \label{eq:QMC-SE-behavior}
  \Sg \propto \frac{1}{\w_n} + \cdots
\end{equation}
as shown in Fig. \ref{fig:fig1} (b). This is consistent with the prediction of MET, see Eq. \eqref{eq:gamma-eq} and \eqref{eq:scaleF-1}.

We analyze the data in two ways. First, we extract the quantum self-energy and compare it to the $T=0$ prediction. To do this we need to remove the thermal part. This is most conveniently done simply by studying the product $\w_n\Sg(\w_n)$.
As discussed in the introduction and the previous Section, according to Eqs.~\eqref{eq:SE-MET} and ~\eqref{eq:Sg-Q-MET-Scaling} we have,
\begin{equation}
  \label{eq:w-Sg}
  \w_n\Sg(\w_n) = \alpha(T) + \gb \wn \sigma(\wn)\left(\frac{\wn}{\w_b}\right)^{2/3}\mathcal{U}\left(\frac{\wn}{\w_b} \right),
\end{equation}
providing we treat $\alpha(T)$ as constant, neglecting its slow frequency dependence, see Eq. \eqref{eq:scaleF-1}. In Eq. \eqref{eq:w-Sg},
$\alpha(T)$ includes both the contribution from $\SigmaT(\wn,T)$,  and corrections from finite size effect (such as a possible small gap due to the mismatch of finite size $h_c$ and the thermodynamic $h_c$). The second part, that is $\wn\SigmaQ(\w_n)$, comes from the MET prediction for $\SigmaQ(\w_n)$, Eq.~\eqref{eq:Sg-Q-MET-Scaling}, which recovers ET prediction Eq. \eqref{eq:sig-ET-scaling} in the low frequency limit ($\wn \ll \w_\text{b}$).

\begin{figure}
  \centering
   \includegraphics[width=0.9\hsize]{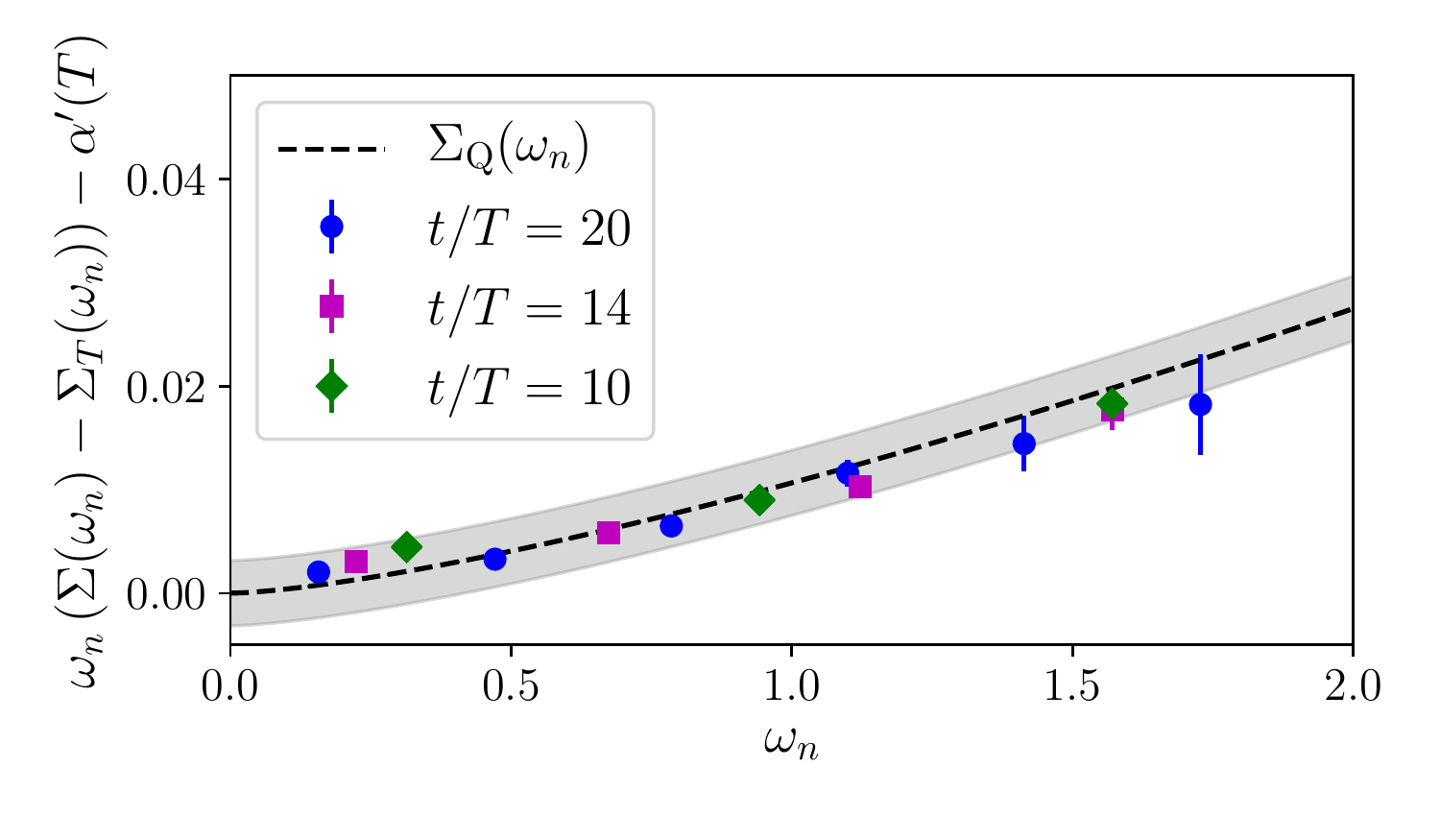}
  
  \caption{\textbf{Extraction of the quantum self energy.} The solid dots correspond to the QMC $\wn\Sg(\wn)$ for different $T$, while for each $T$ dataset, the thermal part and a constant $\alpha'(T)$ has been deducted (see Fig. \ref{fig:comp-qcp-2}). The dashed line corresponds to $\wn\Sigma_\text{Q}(\wn)$, computed at $T=0$, for the bare $\gb$ from the parameters of Sec. \ref{sec:SecII}. The gray shaded area is the 95\% confidence interval.}
  \label{fig:comp-qc-2}
\end{figure}

\begin{figure}
  \centering
  \includegraphics[width=0.9\hsize]{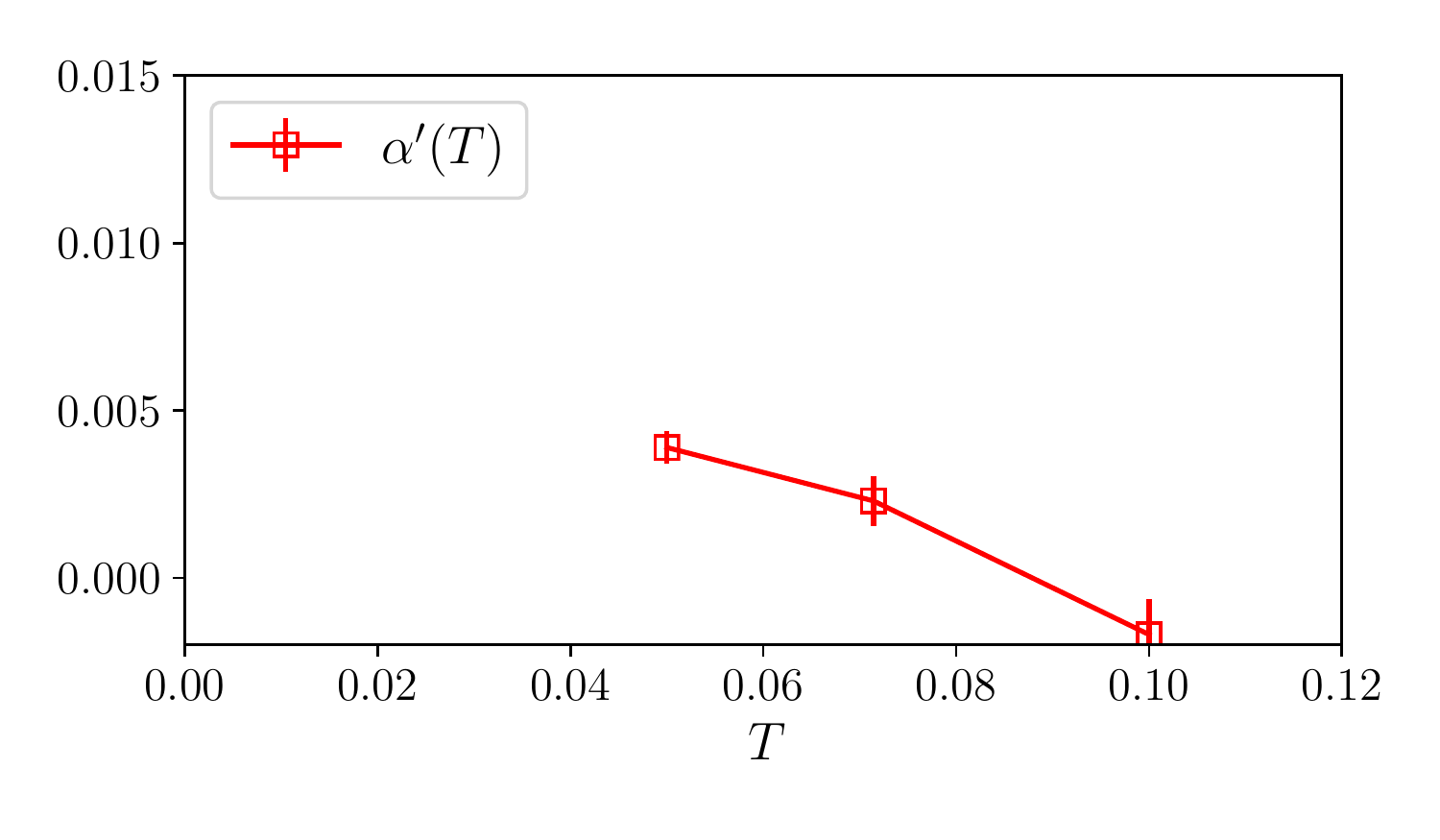}
  \caption{\textbf{The extracted gap contribution to $\wn \Sg(\wn,T)$.} See Eq. \eqref{eq:w-Sg-2} for details.}
  \label{fig:comp-qcp-2}
\end{figure}

We fit Eq. \eqref{eq:w-Sg} to the data for all $T$ simultaneously.
Importantly, in Eq.~\eqref{eq:w-Sg}, the fitting parameters are only the constants $\alpha(T)$ and $\gb$. This is because $\SigmaQ$ is a function only of $\gb$ and system parameters,
see Eq.~\eqref{eq:sig-ET-2}.
Fig. \ref{fig:fig1} (c) from the beginning of our paper depicts the result of our fit. We obtain a fitting of $\gb = 0.245 \pm 0.023$ for 95\% confidence intervals, in excellent agreement with the theory.
Regarding $\alpha(T)$, we find that $\alpha(T) \approx 8 \times 10^{-3}$ is almost a constant, in disagreement with the expected $\propto T$ behavior of $\wn \SigmaT(\wn,T)$. Clearly, part of this discrepancy is due to our neglecting the frequency dependence of $\alpha$. We therefore repeat the analysis using the following fitting procedure,
\begin{equation}
  \label{eq:w-Sg-2}
  \w_n\Sg(\w_n) = \alpha'(T) + \frac{\gb T}{2\pi} S\left(\frac{\upsilon_\text{F} M}{|\wn|}\right) + \gb \wn \sigma(\wn)\left(\frac{\wn}{\wb}\right)^{2/3}\mathcal{U}\left(\frac{\wn}{\wb} \right),
\end{equation}
which takes the full frequency behavior of $\Sg_T$ into account. Guided by the previous fit, we set $\gb = 0.25$ to be the bare one to reduce the number of fitting parameters. We show the result of this fit in Fig. \ref{fig:comp-qc-2} and the extracted $\alpha'(T)$ in Fig. \ref{fig:comp-qcp-2}. The agreement is very good, and we checked that the data collapse can be made even better by allowing $\gb$ to vary somewhat (equivalent  to about 13\% change in the bare vertex $\xi$).
The extracted $\alpha'(T)$ indicates the formation of a small gap forming at around $T=0.1$, which is expected to yield a self-energy contribution of the form $\alpha'(T)/\wn = \Delta^2(T)/\wn$. The gap size $\Delta$ corresponding to $\alpha'(T)$ is much less than the numerical inverse reciprocal lattice spacing, so the appearance of this gap is actually an expected effect, which however is beyond the resolution of the standard methods for veryfing the appearance of long-range order. Thus, our analysis of the self-energy yields a method for more accurately finding the QCP in our system.

\begin{figure}
  \centering
  \includegraphics[width=0.9\hsize]{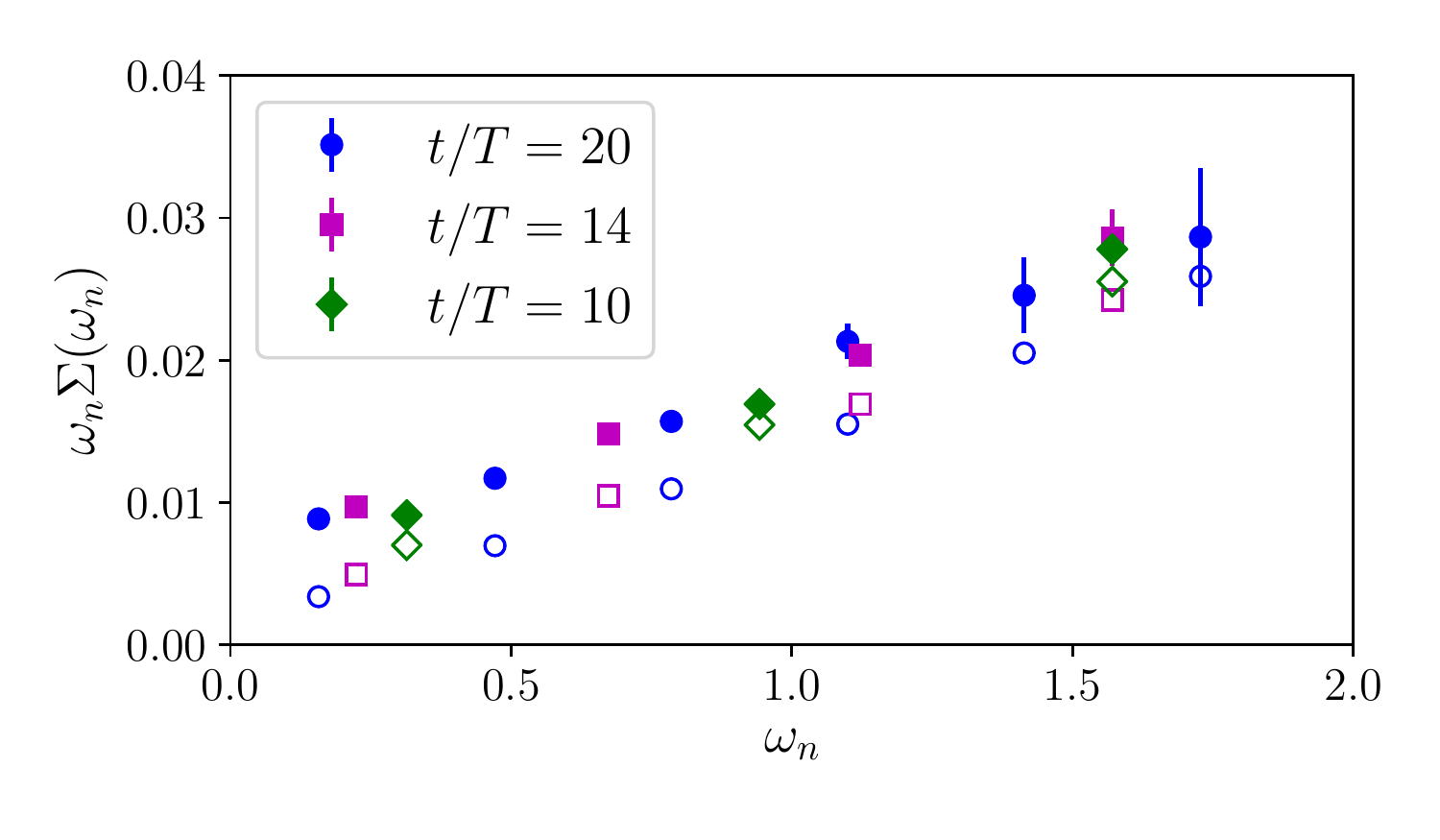}
  \caption{\textbf{Comparison of the full self energy between MET and QMC at the QCP.} The solid dots correspond to the QMC data, and the hollow dots correspond to a numerical summation of the Matsubara sums in Eq.\eqref{eq:sig-low-1}.}
  \label{fig:comp-qcp-3}
\end{figure}
\begin{figure}
  \centering
  \includegraphics[width=0.9\hsize]{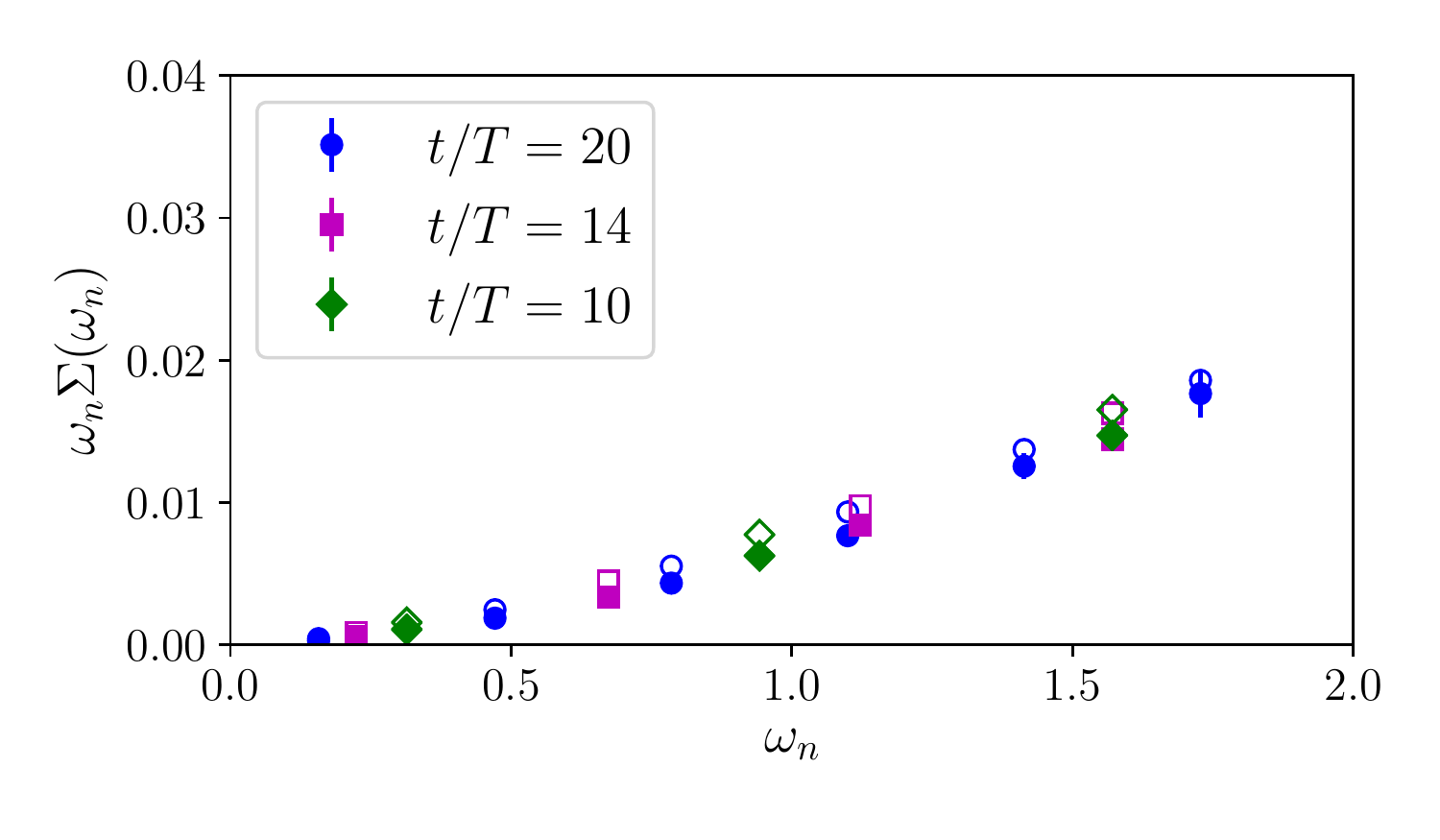}
  \caption{\textbf{Comparison of the full self energy between MET and QMC in the FL region ($h/J=3.6>h_\text{c}/J$).} The solid dots correspond to the QMC data, and the hollow dots correspond to a numerical summation of the Matsubara sums in Eq. \eqref{eq:sig-low-1}.}
  \label{fig:comp-fl-1}
\end{figure}

Here we add a word of caution. Previous work has shown \cite{Chubukov2012,Wang2016,Wu2019} that the first Matsubara frequency does not obey the quantum critical scaling $\SigmaQ(\pi T) \propto (\pi T)^{2/3}$, and therefore should not be included in the fitting procedure. We verified that dropping the first Matsubara point does not change our results.
Also, note that within the error range in Fig. \ref{fig:comp-qc-2} it is possible that $\SigmaQ(\pi T) < 0$.  In fact, it can be verified that $\SigmaQ(\pi T)$ is always negative \cite{DellAnna2006,KleinUn2020}.

To avoid this issue we also numerically computed $\SigmaT(\w_n)$ and $\SigmaQ(\w_n)$ by performing the Matsubara sum in Eq. (\ref{eq:sig-low}), using $\gb = 0.25$. This procedure takes into account the full frequency dependence of $\SigmaT(\w_n)$ as well as finite mass effects and the first Matsubara frequency issues. Fig. \ref{fig:comp-qcp-3} depicts a comparison of the QMC self-energy with the numerical summation. There is an excellent agreement between the two, except for a $T$ dependent constant offset between the MET and QMC results. The result is consistent with the first analysis we performed above.
For completeness, we also performed a comparison between the MET and QMC data for the data in the disordered phase (the FL regime). Fig. \ref{fig:comp-fl-1} shows this comparison, again with very good agreement.

We therefore conclude that we have extracted the quantum self-energy from the QMC data, and that it shows excellent agreement with the expected QC behavior.

\section{Discussion}
\label{sec:SecV}
Non-Fermi liquids play a crucial role in a wide range of quantum many-body phenomena, such as quantum criticality, high-temperature superconductivity in correlated materials, unconventional transport in strange metals, and have been a key focus in the study of modern condensed matter physics~\cite{Hertz1976,Moriya1985,Millis1992,Millis1993,Altshuler1994,Stewart2001,Abanov2003,Metzner2003,Custers2003,Leohneysen2007,Lee2009,Metlitski2010a,Metlitski2010b,Metlitski2010,Lee2018,HQYuanFMQCP2020,Bonesteel1996,Abanov2000,Abanov2004,Rech2006,Abanov2001,Lederer2017,Maslov2010,Fradkin2010,Wang2016,Raghu2015,Metlitski2015,Lederer2015,sachdevbook,Hartnoll2016book,Keimer2015,Berg2019,Xu2019,Chubukov2005,ZHLiu2019}.
Despite of the intensive research efforts, key questions remained open and the problem of NFLs is still one of the most challenging topics in many-body physics, even with the most sophisticated field theoretical treatments~\cite{Abanov2003,Lee2009,Metlitski2010a,Metlitski2010,Lee2018}, powerful numerical many-body algorithms and high-performance supercomputers~\cite{Berg2019,Xu2019,XYXu2017,ZHLiu2019EMUS}. 

Our work
provides a pathway to address a key challenge in the study of non-Fermi liquids, i.e., the
fact that the
smoking-gun signature of non-Fermi liquids (the predicted unconventional low-temperature fermion self-energy), has never been directly observed or verified in large-scale unbiased numerical methods. Though combined numerical and theoretical efforts, we proved that this key signature of non-Fermi liquids can be accessed through QMC simulations, by simply deducting a $\propto 1/\omega_n$ thermal-fluctuation background. This technique enabled us to directly compare numerical results with theoretical predictions, providing a bridge between theoretical, numerical and experimental studies.

Although
this paper mainly focuses on the itinerant ferromagnetism QCP as an example to demonstrate the physics, the technique is universal and can be easily generalized to other itinerant QCPs, such nematic- and AFM-QCPs~\cite{Schattner2016,ZHLiu2018,ZHLiu2019,Bauer2020}. Furthermore, this technique can also be used to explore the predicted nontrivial effects from higher order corrections~\cite{Abanov2003,Lee2009,Metlitski2010a,Metlitski2010,SKLee2017,Lee2018}, and thus open up a pathway towards a full understanding about this challenging subject of  non-Fermi liquids.

\section*{Methods}
\subsection*{Numerical calculations}
\noindent {The numerical results for fermionic and bosonic self-energies  have been obtained using state-of-art determinantal QMC simulations  as reported  in Ref.~\cite{xu2017non}. 
\subsection*{Analytical calculations}
\noindent Analytical calculations have been carried out  diagrammatically within ET and MET, by solving  the set of self-consistent equations for fermionic and bosonic self-energies.

\section*{DATA AVAILABILITY}
\noindent The data that support the findings of this study are available from the first author upon reasonable request.

\section*{Acknowledgements}
We thank Subir Sachdev, Max Metlitski, Yuxuan Wang,
Yoni Schattner, Erez Berg
and Dmitrii Maslov for insightful discussions on fermionic QCPs and NFL. X. Y. Xu also thank Tarun Grover for helpful discussion on related projects. We acknowledge the support from RGC of Hong Kong SAR China through 17303019 and 17301420, MOST through the National Key Research and Development Program (2016YFA0300502). The work by AK and AVC was supported by  the Office of Basic Energy Sciences, U.S. Department of Energy, under award  DE-SC0014402. We thank the Center for Quantum Simulation Sciences in the Institute of Physics, Chinese Academy of Sciences, the Computational Initiative at the Faculty of Science at the University of Hong Kong and the Tianhe platforms at the National Supercomputer Centers in Tianjin and Guangzhou for their technical support and generous allocation of CPU time. This research was initiated at the Aspen Center for Physics, supported by NSF PHY-1066293.

\section*{Competing interests}
\noindent The Authors declare no Competing Financial or Non-Financial Interests.

\section*{AUTHOR CONTRIBUTIONS}
\noindent All authors discussed and designed the study together. X.Y.X. and A.K. analyzed the data. All authors together discussed the theory and drafted the article.

\bibliographystyle{apsrev4-1}
\bibliography{main}

\begin{thebibliography}{77}%
\makeatletter
\providecommand \@ifxundefined [1]{%
 \@ifx{#1\undefined}
}%
\providecommand \@ifnum [1]{%
 \ifnum #1\expandafter \@firstoftwo
 \else \expandafter \@secondoftwo
 \fi
}%
\providecommand \@ifx [1]{%
 \ifx #1\expandafter \@firstoftwo
 \else \expandafter \@secondoftwo
 \fi
}%
\providecommand \natexlab [1]{#1}%
\providecommand \enquote  [1]{``#1''}%
\providecommand \bibnamefont  [1]{#1}%
\providecommand \bibfnamefont [1]{#1}%
\providecommand \citenamefont [1]{#1}%
\providecommand \href@noop [0]{\@secondoftwo}%
\providecommand \href [0]{\begingroup \@sanitize@url \@href}%
\providecommand \@href[1]{\@@startlink{#1}\@@href}%
\providecommand \@@href[1]{\endgroup#1\@@endlink}%
\providecommand \@sanitize@url [0]{\catcode `\\12\catcode `\$12\catcode
  `\&12\catcode `\#12\catcode `\^12\catcode `\_12\catcode `\%12\relax}%
\providecommand \@@startlink[1]{}%
\providecommand \@@endlink[0]{}%
\providecommand \url  [0]{\begingroup\@sanitize@url \@url }%
\providecommand \@url [1]{\endgroup\@href {#1}{\urlprefix }}%
\providecommand \urlprefix  [0]{URL }%
\providecommand \Eprint [0]{\href }%
\providecommand \doibase [0]{http://dx.doi.org/}%
\providecommand \selectlanguage [0]{\@gobble}%
\providecommand \bibinfo  [0]{\@secondoftwo}%
\providecommand \bibfield  [0]{\@secondoftwo}%
\providecommand \translation [1]{[#1]}%
\providecommand \BibitemOpen [0]{}%
\providecommand \bibitemStop [0]{}%
\providecommand \bibitemNoStop [0]{.\EOS\space}%
\providecommand \EOS [0]{\spacefactor3000\relax}%
\providecommand \BibitemShut  [1]{\csname bibitem#1\endcsname}%
\let\auto@bib@innerbib\@empty
\bibitem [{\citenamefont {Sachdev}(2011)}]{sachdevbook}%
  \BibitemOpen
  \bibfield  {author} {\bibinfo {author} {\bibfnamefont {S.}~\bibnamefont
  {Sachdev}},\ }\href {\doibase 10.1017/CBO9780511973765} {\emph {\bibinfo
  {title} {Quantum Phase Transitions}}},\ \bibinfo {edition} {2nd}\ ed.\
  (\bibinfo  {publisher} {Cambridge University Press},\ \bibinfo {year}
  {2011})\BibitemShut {NoStop}%
\bibitem [{\citenamefont {Hertz}(1976)}]{Hertz1976}%
  \BibitemOpen
  \bibfield  {author} {\bibinfo {author} {\bibfnamefont {J.~A.}\ \bibnamefont
  {Hertz}},\ }\href {\doibase 10.1103/PhysRevB.14.1165} {\bibfield  {journal}
  {\bibinfo  {journal} {Phys. Rev. B}\ }\textbf {\bibinfo {volume} {14}},\
  \bibinfo {pages} {1165} (\bibinfo {year} {1976})}\BibitemShut {NoStop}%
\bibitem [{\citenamefont {Moriya}(1985)}]{Moriya1985}%
  \BibitemOpen
  \bibfield  {author} {\bibinfo {author} {\bibfnamefont {T.}~\bibnamefont
  {Moriya}},\ }\href {\doibase 10.1007/978-3-642-82499-9} {\emph {\bibinfo
  {title} {Spin Fluctuations in Itinerant Electron Magnetism}}}\ (\bibinfo
  {publisher} {Springer-Verlag Berlin Heidelberg},\ \bibinfo {year}
  {1985})\BibitemShut {NoStop}%
\bibitem [{\citenamefont {Lee}(1989)}]{Lee1989}%
  \BibitemOpen
  \bibfield  {author} {\bibinfo {author} {\bibfnamefont {P.~A.}\ \bibnamefont
  {Lee}},\ }\href {\doibase 10.1103/PhysRevLett.63.680} {\bibfield  {journal}
  {\bibinfo  {journal} {Phys. Rev. Lett.}\ }\textbf {\bibinfo {volume} {63}},\
  \bibinfo {pages} {680} (\bibinfo {year} {1989})}\BibitemShut {NoStop}%
\bibitem [{\citenamefont {Millis}(1992)}]{Millis1992}%
  \BibitemOpen
  \bibfield  {author} {\bibinfo {author} {\bibfnamefont {A.~J.}\ \bibnamefont
  {Millis}},\ }\href {\doibase 10.1103/PhysRevB.45.13047} {\bibfield  {journal}
  {\bibinfo  {journal} {Phys. Rev. B}\ }\textbf {\bibinfo {volume} {45}},\
  \bibinfo {pages} {13047} (\bibinfo {year} {1992})}\BibitemShut {NoStop}%
\bibitem [{\citenamefont {Millis}(1993)}]{Millis1993}%
  \BibitemOpen
  \bibfield  {author} {\bibinfo {author} {\bibfnamefont {A.~J.}\ \bibnamefont
  {Millis}},\ }\href {\doibase 10.1103/PhysRevB.48.7183} {\bibfield  {journal}
  {\bibinfo  {journal} {Phys. Rev. B}\ }\textbf {\bibinfo {volume} {48}},\
  \bibinfo {pages} {7183} (\bibinfo {year} {1993})}\BibitemShut {NoStop}%
\bibitem [{\citenamefont {Altshuler}\ \emph {et~al.}(1994)\citenamefont
  {Altshuler}, \citenamefont {Ioffe},\ and\ \citenamefont
  {Millis}}]{Altshuler1994}%
  \BibitemOpen
  \bibfield  {author} {\bibinfo {author} {\bibfnamefont {B.~L.}\ \bibnamefont
  {Altshuler}}, \bibinfo {author} {\bibfnamefont {L.~B.}\ \bibnamefont
  {Ioffe}}, \ and\ \bibinfo {author} {\bibfnamefont {A.~J.}\ \bibnamefont
  {Millis}},\ }\href {\doibase 10.1103/PhysRevB.50.14048} {\bibfield  {journal}
  {\bibinfo  {journal} {Phys. Rev. B}\ }\textbf {\bibinfo {volume} {50}},\
  \bibinfo {pages} {14048} (\bibinfo {year} {1994})}\BibitemShut {NoStop}%
\bibitem [{\citenamefont {Polchinski}(1994)}]{Polchinski1994}%
  \BibitemOpen
  \bibfield  {author} {\bibinfo {author} {\bibfnamefont {J.}~\bibnamefont
  {Polchinski}},\ }\href {\doibase
  https://doi.org/10.1016/0550-3213(94)90449-9} {\bibfield  {journal} {\bibinfo
   {journal} {Nuclear Physics B}\ }\textbf {\bibinfo {volume} {422}},\ \bibinfo
  {pages} {617 } (\bibinfo {year} {1994})}\BibitemShut {NoStop}%
\bibitem [{\citenamefont {Nayak}\ and\ \citenamefont
  {Wilczek}(1994)}]{Nayak1994}%
  \BibitemOpen
  \bibfield  {author} {\bibinfo {author} {\bibfnamefont {C.}~\bibnamefont
  {Nayak}}\ and\ \bibinfo {author} {\bibfnamefont {F.}~\bibnamefont
  {Wilczek}},\ }\href
  {http://www.sciencedirect.com/science/article/pii/0550321394904774}
  {\bibfield  {journal} {\bibinfo  {journal} {Nuclear Physics B}\ }\textbf
  {\bibinfo {volume} {417}},\ \bibinfo {pages} {359} (\bibinfo {year}
  {1994})}\BibitemShut {NoStop}%
\bibitem [{\citenamefont {Son}(1999)}]{son}%
  \BibitemOpen
  \bibfield  {author} {\bibinfo {author} {\bibfnamefont {D.~T.}\ \bibnamefont
  {Son}},\ }\href {\doibase 10.1103/PhysRevD.59.094019} {\bibfield  {journal}
  {\bibinfo  {journal} {Phys. Rev. D}\ }\textbf {\bibinfo {volume} {59}},\
  \bibinfo {pages} {094019} (\bibinfo {year} {1999})}\BibitemShut {NoStop}%
\bibitem [{\citenamefont {Chubukov}\ and\ \citenamefont
  {Schmalian}(2005)}]{son2}%
  \BibitemOpen
  \bibfield  {author} {\bibinfo {author} {\bibfnamefont {A.~V.}\ \bibnamefont
  {Chubukov}}\ and\ \bibinfo {author} {\bibfnamefont {J.}~\bibnamefont
  {Schmalian}},\ }\href {\doibase 10.1103/PhysRevB.72.174520} {\bibfield
  {journal} {\bibinfo  {journal} {Phys. Rev. B}\ }\textbf {\bibinfo {volume}
  {72}},\ \bibinfo {pages} {174520} (\bibinfo {year} {2005})}\BibitemShut
  {NoStop}%
\bibitem [{\citenamefont {Abanov}\ and\ \citenamefont
  {Chubukov}(2000)}]{Abanov2000}%
  \BibitemOpen
  \bibfield  {author} {\bibinfo {author} {\bibfnamefont {A.}~\bibnamefont
  {Abanov}}\ and\ \bibinfo {author} {\bibfnamefont {A.~V.}\ \bibnamefont
  {Chubukov}},\ }\href {\doibase 10.1103/PhysRevLett.84.5608} {\bibfield
  {journal} {\bibinfo  {journal} {Phys. Rev. Lett.}\ }\textbf {\bibinfo
  {volume} {84}},\ \bibinfo {pages} {5608} (\bibinfo {year}
  {2000})}\BibitemShut {NoStop}%
\bibitem [{\citenamefont {Oganesyan}\ \emph {et~al.}(2001)\citenamefont
  {Oganesyan}, \citenamefont {Kivelson},\ and\ \citenamefont
  {Fradkin}}]{Oganesyan2001}%
  \BibitemOpen
  \bibfield  {author} {\bibinfo {author} {\bibfnamefont {V.}~\bibnamefont
  {Oganesyan}}, \bibinfo {author} {\bibfnamefont {S.~A.}\ \bibnamefont
  {Kivelson}}, \ and\ \bibinfo {author} {\bibfnamefont {E.}~\bibnamefont
  {Fradkin}},\ }\href {\doibase 10.1103/PhysRevB.64.195109} {\bibfield
  {journal} {\bibinfo  {journal} {Phys. Rev. B}\ }\textbf {\bibinfo {volume}
  {64}},\ \bibinfo {pages} {195109} (\bibinfo {year} {2001})}\BibitemShut
  {NoStop}%
\bibitem [{\citenamefont {Abanov}\ \emph {et~al.}(2001)\citenamefont {Abanov},
  \citenamefont {Chubukov},\ and\ \citenamefont {Finkel'stein}}]{Abanov2001}%
  \BibitemOpen
  \bibfield  {author} {\bibinfo {author} {\bibfnamefont {A.}~\bibnamefont
  {Abanov}}, \bibinfo {author} {\bibfnamefont {A.~V.}\ \bibnamefont
  {Chubukov}}, \ and\ \bibinfo {author} {\bibfnamefont {A.~M.}\ \bibnamefont
  {Finkel'stein}},\ }\href {http://stacks.iop.org/0295-5075/54/i=4/a=488}
  {\bibfield  {journal} {\bibinfo  {journal} {EPL (Europhysics Letters)}\
  }\textbf {\bibinfo {volume} {54}},\ \bibinfo {pages} {488} (\bibinfo {year}
  {2001})}\BibitemShut {NoStop}%
\bibitem [{\citenamefont {Stewart}(2001)}]{Stewart2001}%
  \BibitemOpen
  \bibfield  {author} {\bibinfo {author} {\bibfnamefont {G.~R.}\ \bibnamefont
  {Stewart}},\ }\href {\doibase 10.1103/RevModPhys.73.797} {\bibfield
  {journal} {\bibinfo  {journal} {Rev. Mod. Phys.}\ }\textbf {\bibinfo {volume}
  {73}},\ \bibinfo {pages} {797} (\bibinfo {year} {2001})}\BibitemShut
  {NoStop}%
\bibitem [{\citenamefont {Abanov}\ \emph {et~al.}(2003)\citenamefont {Abanov},
  \citenamefont {Chubukov},\ and\ \citenamefont {Schmalian}}]{Abanov2003}%
  \BibitemOpen
  \bibfield  {author} {\bibinfo {author} {\bibfnamefont {A.}~\bibnamefont
  {Abanov}}, \bibinfo {author} {\bibfnamefont {A.~V.}\ \bibnamefont
  {Chubukov}}, \ and\ \bibinfo {author} {\bibfnamefont {J.}~\bibnamefont
  {Schmalian}},\ }\bibfield  {booktitle} {\emph {\bibinfo {booktitle} {Advances
  in Physics}},\ }\href {\doibase 10.1080/0001873021000057123} {\bibfield
  {journal} {\bibinfo  {journal} {Advances in Physics}\ }\textbf {\bibinfo
  {volume} {52}},\ \bibinfo {pages} {119} (\bibinfo {year} {2003})}\BibitemShut
  {NoStop}%
\bibitem [{\citenamefont {Metzner}\ \emph {et~al.}(2003)\citenamefont
  {Metzner}, \citenamefont {Rohe},\ and\ \citenamefont
  {Andergassen}}]{Metzner2003}%
  \BibitemOpen
  \bibfield  {author} {\bibinfo {author} {\bibfnamefont {W.}~\bibnamefont
  {Metzner}}, \bibinfo {author} {\bibfnamefont {D.}~\bibnamefont {Rohe}}, \
  and\ \bibinfo {author} {\bibfnamefont {S.}~\bibnamefont {Andergassen}},\
  }\href {\doibase 10.1103/PhysRevLett.91.066402} {\bibfield  {journal}
  {\bibinfo  {journal} {Phys. Rev. Lett.}\ }\textbf {\bibinfo {volume} {91}},\
  \bibinfo {pages} {066402} (\bibinfo {year} {2003})}\BibitemShut {NoStop}%
\bibitem [{\citenamefont {Custers}\ \emph {et~al.}(2003)\citenamefont
  {Custers}, \citenamefont {Gegenwart}, \citenamefont {Wilhelm}, \citenamefont
  {Neumaier}, \citenamefont {Tokiwa}, \citenamefont {Trovarelli}, \citenamefont
  {Geibel}, \citenamefont {Steglich}, \citenamefont {Pépin},\ and\
  \citenamefont {Coleman}}]{Custers2003}%
  \BibitemOpen
  \bibfield  {author} {\bibinfo {author} {\bibfnamefont {J.}~\bibnamefont
  {Custers}}, \bibinfo {author} {\bibfnamefont {P.}~\bibnamefont {Gegenwart}},
  \bibinfo {author} {\bibfnamefont {H.}~\bibnamefont {Wilhelm}}, \bibinfo
  {author} {\bibfnamefont {K.}~\bibnamefont {Neumaier}}, \bibinfo {author}
  {\bibfnamefont {Y.}~\bibnamefont {Tokiwa}}, \bibinfo {author} {\bibfnamefont
  {O.}~\bibnamefont {Trovarelli}}, \bibinfo {author} {\bibfnamefont
  {C.}~\bibnamefont {Geibel}}, \bibinfo {author} {\bibfnamefont
  {F.}~\bibnamefont {Steglich}}, \bibinfo {author} {\bibfnamefont
  {C.}~\bibnamefont {Pépin}}, \ and\ \bibinfo {author} {\bibfnamefont
  {P.}~\bibnamefont {Coleman}},\ }\href {\doibase 10.1038/nature01774}
  {\bibfield  {journal} {\bibinfo  {journal} {Nature}\ }\textbf {\bibinfo
  {volume} {424}},\ \bibinfo {pages} {524} (\bibinfo {year}
  {2003})}\BibitemShut {NoStop}%
\bibitem [{\citenamefont {Abanov}\ and\ \citenamefont
  {Chubukov}(2004)}]{Abanov2004}%
  \BibitemOpen
  \bibfield  {author} {\bibinfo {author} {\bibfnamefont {A.}~\bibnamefont
  {Abanov}}\ and\ \bibinfo {author} {\bibfnamefont {A.}~\bibnamefont
  {Chubukov}},\ }\href {\doibase 10.1103/PhysRevLett.93.255702} {\bibfield
  {journal} {\bibinfo  {journal} {Phys. Rev. Lett.}\ }\textbf {\bibinfo
  {volume} {93}},\ \bibinfo {pages} {255702} (\bibinfo {year}
  {2004})}\BibitemShut {NoStop}%
\bibitem [{\citenamefont {Chubukov}(2005{\natexlab{a}})}]{Chubukov2005a}%
  \BibitemOpen
  \bibfield  {author} {\bibinfo {author} {\bibfnamefont {A.~V.}\ \bibnamefont
  {Chubukov}},\ }\href {\doibase 10.1103/PhysRevB.71.245123} {\bibfield
  {journal} {\bibinfo  {journal} {Phys. Rev. B}\ }\textbf {\bibinfo {volume}
  {71}},\ \bibinfo {pages} {245123} (\bibinfo {year}
  {2005}{\natexlab{a}})}\BibitemShut {NoStop}%
\bibitem [{\citenamefont {Dell'Anna}\ and\ \citenamefont
  {Metzner}(2006)}]{DellAnna2006}%
  \BibitemOpen
  \bibfield  {author} {\bibinfo {author} {\bibfnamefont {L.}~\bibnamefont
  {Dell'Anna}}\ and\ \bibinfo {author} {\bibfnamefont {W.}~\bibnamefont
  {Metzner}},\ }\href {\doibase 10.1103/PhysRevB.73.045127} {\bibfield
  {journal} {\bibinfo  {journal} {Phys. Rev. B}\ }\textbf {\bibinfo {volume}
  {73}},\ \bibinfo {pages} {045127} (\bibinfo {year} {2006})}\BibitemShut
  {NoStop}%
\bibitem [{\citenamefont {Rech}\ \emph {et~al.}(2006)\citenamefont {Rech},
  \citenamefont {P\'epin},\ and\ \citenamefont {Chubukov}}]{Rech2006}%
  \BibitemOpen
  \bibfield  {author} {\bibinfo {author} {\bibfnamefont {J.}~\bibnamefont
  {Rech}}, \bibinfo {author} {\bibfnamefont {C.}~\bibnamefont {P\'epin}}, \
  and\ \bibinfo {author} {\bibfnamefont {A.~V.}\ \bibnamefont {Chubukov}},\
  }\href {\doibase 10.1103/PhysRevB.74.195126} {\bibfield  {journal} {\bibinfo
  {journal} {Phys. Rev. B}\ }\textbf {\bibinfo {volume} {74}},\ \bibinfo
  {pages} {195126} (\bibinfo {year} {2006})}\BibitemShut {NoStop}%
\bibitem [{\citenamefont {Maslov}\ \emph {et~al.}(2006)\citenamefont {Maslov},
  \citenamefont {Chubukov},\ and\ \citenamefont {Saha}}]{Maslov2006}%
  \BibitemOpen
  \bibfield  {author} {\bibinfo {author} {\bibfnamefont {D.~L.}\ \bibnamefont
  {Maslov}}, \bibinfo {author} {\bibfnamefont {A.~V.}\ \bibnamefont
  {Chubukov}}, \ and\ \bibinfo {author} {\bibfnamefont {R.}~\bibnamefont
  {Saha}},\ }\href {\doibase 10.1103/PhysRevB.74.220402} {\bibfield  {journal}
  {\bibinfo  {journal} {Phys. Rev. B}\ }\textbf {\bibinfo {volume} {74}},\
  \bibinfo {pages} {220402} (\bibinfo {year} {2006})}\BibitemShut {NoStop}%
\bibitem [{\citenamefont {L\"ohneysen}\ \emph {et~al.}(2007)\citenamefont
  {L\"ohneysen}, \citenamefont {Rosch}, \citenamefont {Vojta},\ and\
  \citenamefont {W\"olfle}}]{Leohneysen2007}%
  \BibitemOpen
  \bibfield  {author} {\bibinfo {author} {\bibfnamefont {H.~v.}\ \bibnamefont
  {L\"ohneysen}}, \bibinfo {author} {\bibfnamefont {A.}~\bibnamefont {Rosch}},
  \bibinfo {author} {\bibfnamefont {M.}~\bibnamefont {Vojta}}, \ and\ \bibinfo
  {author} {\bibfnamefont {P.}~\bibnamefont {W\"olfle}},\ }\href {\doibase
  10.1103/RevModPhys.79.1015} {\bibfield  {journal} {\bibinfo  {journal} {Rev.
  Mod. Phys.}\ }\textbf {\bibinfo {volume} {79}},\ \bibinfo {pages} {1015}
  (\bibinfo {year} {2007})}\BibitemShut {NoStop}%
\bibitem [{\citenamefont {Lee}(2009)}]{Lee2009}%
  \BibitemOpen
  \bibfield  {author} {\bibinfo {author} {\bibfnamefont {S.-S.}\ \bibnamefont
  {Lee}},\ }\href {\doibase 10.1103/PhysRevB.80.165102} {\bibfield  {journal}
  {\bibinfo  {journal} {Phys. Rev. B}\ }\textbf {\bibinfo {volume} {80}},\
  \bibinfo {pages} {165102} (\bibinfo {year} {2009})}\BibitemShut {NoStop}%
\bibitem [{\citenamefont {Maslov}\ and\ \citenamefont
  {Chubukov}(2009)}]{Maslov2009}%
  \BibitemOpen
  \bibfield  {author} {\bibinfo {author} {\bibfnamefont {D.~L.}\ \bibnamefont
  {Maslov}}\ and\ \bibinfo {author} {\bibfnamefont {A.~V.}\ \bibnamefont
  {Chubukov}},\ }\href {\doibase 10.1103/PhysRevB.79.075112} {\bibfield
  {journal} {\bibinfo  {journal} {Phys. Rev. B}\ }\textbf {\bibinfo {volume}
  {79}},\ \bibinfo {pages} {075112} (\bibinfo {year} {2009})}\BibitemShut
  {NoStop}%
\bibitem [{\citenamefont {Metlitski}\ and\ \citenamefont
  {Sachdev}(2010{\natexlab{a}})}]{Metlitski2010a}%
  \BibitemOpen
  \bibfield  {author} {\bibinfo {author} {\bibfnamefont {M.~A.}\ \bibnamefont
  {Metlitski}}\ and\ \bibinfo {author} {\bibfnamefont {S.}~\bibnamefont
  {Sachdev}},\ }\href {\doibase 10.1103/PhysRevB.82.075127} {\bibfield
  {journal} {\bibinfo  {journal} {Phys. Rev. B}\ }\textbf {\bibinfo {volume}
  {82}},\ \bibinfo {pages} {075127} (\bibinfo {year}
  {2010}{\natexlab{a}})}\BibitemShut {NoStop}%
\bibitem [{\citenamefont {Metlitski}\ and\ \citenamefont
  {Sachdev}(2010{\natexlab{b}})}]{Metlitski2010b}%
  \BibitemOpen
  \bibfield  {author} {\bibinfo {author} {\bibfnamefont {M.~A.}\ \bibnamefont
  {Metlitski}}\ and\ \bibinfo {author} {\bibfnamefont {S.}~\bibnamefont
  {Sachdev}},\ }\href {http://stacks.iop.org/1367-2630/12/i=10/a=105007}
  {\bibfield  {journal} {\bibinfo  {journal} {New Journal of Physics}\ }\textbf
  {\bibinfo {volume} {12}},\ \bibinfo {pages} {105007} (\bibinfo {year}
  {2010}{\natexlab{b}})}\BibitemShut {NoStop}%
\bibitem [{\citenamefont {Metlitski}\ and\ \citenamefont
  {Sachdev}(2010{\natexlab{c}})}]{Metlitski2010}%
  \BibitemOpen
  \bibfield  {author} {\bibinfo {author} {\bibfnamefont {M.~A.}\ \bibnamefont
  {Metlitski}}\ and\ \bibinfo {author} {\bibfnamefont {S.}~\bibnamefont
  {Sachdev}},\ }\href {\doibase 10.1103/PhysRevB.82.075128} {\bibfield
  {journal} {\bibinfo  {journal} {Phys. Rev. B}\ }\textbf {\bibinfo {volume}
  {82}},\ \bibinfo {pages} {075128} (\bibinfo {year}
  {2010}{\natexlab{c}})}\BibitemShut {NoStop}%
\bibitem [{\citenamefont {Mross}\ \emph {et~al.}(2010)\citenamefont {Mross},
  \citenamefont {McGreevy}, \citenamefont {Liu},\ and\ \citenamefont
  {Senthil}}]{Mross2010}%
  \BibitemOpen
  \bibfield  {author} {\bibinfo {author} {\bibfnamefont {D.~F.}\ \bibnamefont
  {Mross}}, \bibinfo {author} {\bibfnamefont {J.}~\bibnamefont {McGreevy}},
  \bibinfo {author} {\bibfnamefont {H.}~\bibnamefont {Liu}}, \ and\ \bibinfo
  {author} {\bibfnamefont {T.}~\bibnamefont {Senthil}},\ }\href {\doibase
  10.1103/PhysRevB.82.045121} {\bibfield  {journal} {\bibinfo  {journal} {Phys.
  Rev. B}\ }\textbf {\bibinfo {volume} {82}},\ \bibinfo {pages} {045121}
  (\bibinfo {year} {2010})}\BibitemShut {NoStop}%
\bibitem [{\citenamefont {Holder}\ and\ \citenamefont
  {Metzner}(2015{\natexlab{a}})}]{Holder2015}%
  \BibitemOpen
  \bibfield  {author} {\bibinfo {author} {\bibfnamefont {T.}~\bibnamefont
  {Holder}}\ and\ \bibinfo {author} {\bibfnamefont {W.}~\bibnamefont
  {Metzner}},\ }\href {\doibase 10.1103/PhysRevB.92.245128} {\bibfield
  {journal} {\bibinfo  {journal} {Phys. Rev. B}\ }\textbf {\bibinfo {volume}
  {92}},\ \bibinfo {pages} {245128} (\bibinfo {year}
  {2015}{\natexlab{a}})}\BibitemShut {NoStop}%
\bibitem [{\citenamefont {Holder}\ and\ \citenamefont
  {Metzner}(2015{\natexlab{b}})}]{Holder2015a}%
  \BibitemOpen
  \bibfield  {author} {\bibinfo {author} {\bibfnamefont {T.}~\bibnamefont
  {Holder}}\ and\ \bibinfo {author} {\bibfnamefont {W.}~\bibnamefont
  {Metzner}},\ }\href {\doibase 10.1103/PhysRevB.92.041112} {\bibfield
  {journal} {\bibinfo  {journal} {Phys. Rev. B}\ }\textbf {\bibinfo {volume}
  {92}},\ \bibinfo {pages} {041112} (\bibinfo {year}
  {2015}{\natexlab{b}})}\BibitemShut {NoStop}%
\bibitem [{\citenamefont {Wang}\ \emph {et~al.}(2016)\citenamefont {Wang},
  \citenamefont {Abanov}, \citenamefont {Altshuler}, \citenamefont
  {Yuzbashyan},\ and\ \citenamefont {Chubukov}}]{Wang2016}%
  \BibitemOpen
  \bibfield  {author} {\bibinfo {author} {\bibfnamefont {Y.}~\bibnamefont
  {Wang}}, \bibinfo {author} {\bibfnamefont {A.}~\bibnamefont {Abanov}},
  \bibinfo {author} {\bibfnamefont {B.~L.}\ \bibnamefont {Altshuler}}, \bibinfo
  {author} {\bibfnamefont {E.~A.}\ \bibnamefont {Yuzbashyan}}, \ and\ \bibinfo
  {author} {\bibfnamefont {A.~V.}\ \bibnamefont {Chubukov}},\ }\href {\doibase
  10.1103/PhysRevLett.117.157001} {\bibfield  {journal} {\bibinfo  {journal}
  {Phys. Rev. Lett.}\ }\textbf {\bibinfo {volume} {117}},\ \bibinfo {pages}
  {157001} (\bibinfo {year} {2016})}\BibitemShut {NoStop}%
\bibitem [{\citenamefont {Wang}\ and\ \citenamefont
  {Torroba}(2017)}]{Wang2017a}%
  \BibitemOpen
  \bibfield  {author} {\bibinfo {author} {\bibfnamefont {H.}~\bibnamefont
  {Wang}}\ and\ \bibinfo {author} {\bibfnamefont {G.}~\bibnamefont {Torroba}},\
  }\href {\doibase 10.1103/PhysRevB.96.144508} {\bibfield  {journal} {\bibinfo
  {journal} {Phys. Rev. B}\ }\textbf {\bibinfo {volume} {96}},\ \bibinfo
  {pages} {144508} (\bibinfo {year} {2017})}\BibitemShut {NoStop}%
\bibitem [{\citenamefont {Lee}(2018)}]{Lee2018}%
  \BibitemOpen
  \bibfield  {author} {\bibinfo {author} {\bibfnamefont {S.-S.}\ \bibnamefont
  {Lee}},\ }\bibfield  {booktitle} {\emph {\bibinfo {booktitle} {Annual Review
  of Condensed Matter Physics}},\ }\href {\doibase
  10.1146/annurev-conmatphys-031016-025531} {\bibfield  {journal} {\bibinfo
  {journal} {Annu. Rev. Condens. Matter Phys.}\ }\textbf {\bibinfo {volume}
  {9}},\ \bibinfo {pages} {227} (\bibinfo {year} {2018})}\BibitemShut {NoStop}%
\bibitem [{\citenamefont {Shen}\ \emph {et~al.}(2020)\citenamefont {Shen},
  \citenamefont {Zhang}, \citenamefont {Komijani}, \citenamefont {Nicklas},
  \citenamefont {Borth}, \citenamefont {Wang}, \citenamefont {Chen},
  \citenamefont {Nie}, \citenamefont {Li}, \citenamefont {Lu}, \citenamefont
  {Lee}, \citenamefont {Smidman}, \citenamefont {Steglich}, \citenamefont
  {Coleman},\ and\ \citenamefont {Yuan}}]{HQYuanFMQCP2020}%
  \BibitemOpen
  \bibfield  {author} {\bibinfo {author} {\bibfnamefont {B.}~\bibnamefont
  {Shen}}, \bibinfo {author} {\bibfnamefont {Y.}~\bibnamefont {Zhang}},
  \bibinfo {author} {\bibfnamefont {Y.}~\bibnamefont {Komijani}}, \bibinfo
  {author} {\bibfnamefont {M.}~\bibnamefont {Nicklas}}, \bibinfo {author}
  {\bibfnamefont {R.}~\bibnamefont {Borth}}, \bibinfo {author} {\bibfnamefont
  {A.}~\bibnamefont {Wang}}, \bibinfo {author} {\bibfnamefont {Y.}~\bibnamefont
  {Chen}}, \bibinfo {author} {\bibfnamefont {Z.}~\bibnamefont {Nie}}, \bibinfo
  {author} {\bibfnamefont {R.}~\bibnamefont {Li}}, \bibinfo {author}
  {\bibfnamefont {X.}~\bibnamefont {Lu}}, \bibinfo {author} {\bibfnamefont
  {H.}~\bibnamefont {Lee}}, \bibinfo {author} {\bibfnamefont {M.}~\bibnamefont
  {Smidman}}, \bibinfo {author} {\bibfnamefont {F.}~\bibnamefont {Steglich}},
  \bibinfo {author} {\bibfnamefont {P.}~\bibnamefont {Coleman}}, \ and\
  \bibinfo {author} {\bibfnamefont {H.}~\bibnamefont {Yuan}},\ }\href {\doibase
  10.1038/s41586-020-2052-z} {\bibfield  {journal} {\bibinfo  {journal}
  {Nature}\ }\textbf {\bibinfo {volume} {579}},\ \bibinfo {pages} {51}
  (\bibinfo {year} {2020})}\BibitemShut {NoStop}%
\bibitem [{\citenamefont {Damia}\ \emph {et~al.}(2019)\citenamefont {Damia},
  \citenamefont {Kachru}, \citenamefont {Raghu},\ and\ \citenamefont
  {Torroba}}]{Torroba2019}%
  \BibitemOpen
  \bibfield  {author} {\bibinfo {author} {\bibfnamefont {J.~A.}\ \bibnamefont
  {Damia}}, \bibinfo {author} {\bibfnamefont {S.}~\bibnamefont {Kachru}},
  \bibinfo {author} {\bibfnamefont {S.}~\bibnamefont {Raghu}}, \ and\ \bibinfo
  {author} {\bibfnamefont {G.}~\bibnamefont {Torroba}},\ }\href {\doibase
  10.1103/PhysRevLett.123.096402} {\bibfield  {journal} {\bibinfo  {journal}
  {Phys. Rev. Lett.}\ }\textbf {\bibinfo {volume} {123}},\ \bibinfo {pages}
  {096402} (\bibinfo {year} {2019})}\BibitemShut {NoStop}%
\bibitem [{\citenamefont {Wu}\ \emph {et~al.}(2019)\citenamefont {Wu},
  \citenamefont {Abanov}, \citenamefont {Wang},\ and\ \citenamefont
  {Chubukov}}]{Wu2019}%
  \BibitemOpen
  \bibfield  {author} {\bibinfo {author} {\bibfnamefont {Y.-M.}\ \bibnamefont
  {Wu}}, \bibinfo {author} {\bibfnamefont {A.}~\bibnamefont {Abanov}}, \bibinfo
  {author} {\bibfnamefont {Y.}~\bibnamefont {Wang}}, \ and\ \bibinfo {author}
  {\bibfnamefont {A.~V.}\ \bibnamefont {Chubukov}},\ }\href {\doibase
  10.1103/PhysRevB.99.144512} {\bibfield  {journal} {\bibinfo  {journal} {Phys.
  Rev. B}\ }\textbf {\bibinfo {volume} {99}},\ \bibinfo {pages} {144512}
  (\bibinfo {year} {2019})}\BibitemShut {NoStop}%
\bibitem [{\citenamefont {Esterlis}\ and\ \citenamefont
  {Schmalian}(2019)}]{Esterlis2019}%
  \BibitemOpen
  \bibfield  {author} {\bibinfo {author} {\bibfnamefont {I.}~\bibnamefont
  {Esterlis}}\ and\ \bibinfo {author} {\bibfnamefont {J.}~\bibnamefont
  {Schmalian}},\ }\href {\doibase 10.1103/PhysRevB.100.115132} {\bibfield
  {journal} {\bibinfo  {journal} {Phys. Rev. B}\ }\textbf {\bibinfo {volume}
  {100}},\ \bibinfo {pages} {115132} (\bibinfo {year} {2019})}\BibitemShut
  {NoStop}%
\bibitem [{\citenamefont {Wang}(2020)}]{Wang2020}%
  \BibitemOpen
  \bibfield  {author} {\bibinfo {author} {\bibfnamefont {Y.}~\bibnamefont
  {Wang}},\ }\href {\doibase 10.1103/PhysRevLett.124.017002} {\bibfield
  {journal} {\bibinfo  {journal} {Phys. Rev. Lett.}\ }\textbf {\bibinfo
  {volume} {124}},\ \bibinfo {pages} {017002} (\bibinfo {year}
  {2020})}\BibitemShut {NoStop}%
\bibitem [{\citenamefont {{Pan}}\ \emph {et~al.}(2020)\citenamefont {{Pan}},
  \citenamefont {{Wang}},\ and\ \citenamefont {{Meng}}}]{GPPan2020}%
  \BibitemOpen
  \bibfield  {author} {\bibinfo {author} {\bibfnamefont {G.}~\bibnamefont
  {{Pan}}}, \bibinfo {author} {\bibfnamefont {Y.}~\bibnamefont {{Wang}}}, \
  and\ \bibinfo {author} {\bibfnamefont {Z.~Y.}\ \bibnamefont {{Meng}}},\
  }\href@noop {} {\bibfield  {journal} {\bibinfo  {journal} {arXiv e-prints}\
  ,\ \bibinfo {pages} {arXiv:2001.06586}} (\bibinfo {year} {2020})},\ \Eprint
  {http://arxiv.org/abs/2001.06586} {arXiv:2001.06586 [cond-mat.str-el]}
  \BibitemShut {NoStop}%
\bibitem [{\citenamefont {Xu}\ \emph {et~al.}(2020)\citenamefont {Xu},
  \citenamefont {Geng}, \citenamefont {Wu}, \citenamefont {Jian},\ and\
  \citenamefont {Xu}}]{YichenXu2020}%
  \BibitemOpen
  \bibfield  {author} {\bibinfo {author} {\bibfnamefont {Y.}~\bibnamefont
  {Xu}}, \bibinfo {author} {\bibfnamefont {H.}~\bibnamefont {Geng}}, \bibinfo
  {author} {\bibfnamefont {X.-C.}\ \bibnamefont {Wu}}, \bibinfo {author}
  {\bibfnamefont {C.-M.}\ \bibnamefont {Jian}}, \ and\ \bibinfo {author}
  {\bibfnamefont {C.}~\bibnamefont {Xu}},\ }\href {\doibase
  10.1088/1742-5468/ab99a0} {\bibfield  {journal} {\bibinfo  {journal} {Journal
  of Statistical Mechanics: Theory and Experiment}\ }\textbf {\bibinfo {volume}
  {2020}},\ \bibinfo {pages} {073102} (\bibinfo {year} {2020})}\BibitemShut
  {NoStop}%
\bibitem [{\citenamefont {{Aguilera Damia}}\ \emph {et~al.}(2020)\citenamefont
  {{Aguilera Damia}}, \citenamefont {{Solis}},\ and\ \citenamefont
  {{Torroba}}}]{Damia2020}%
  \BibitemOpen
  \bibfield  {author} {\bibinfo {author} {\bibfnamefont {J.}~\bibnamefont
  {{Aguilera Damia}}}, \bibinfo {author} {\bibfnamefont {M.}~\bibnamefont
  {{Solis}}}, \ and\ \bibinfo {author} {\bibfnamefont {G.}~\bibnamefont
  {{Torroba}}},\ }\href@noop {} {\bibfield  {journal} {\bibinfo  {journal}
  {arXiv e-prints}\ ,\ \bibinfo {eid} {arXiv:2004.05181}} (\bibinfo {year}
  {2020})},\ \Eprint {http://arxiv.org/abs/2004.05181} {arXiv:2004.05181
  [cond-mat.str-el]} \BibitemShut {NoStop}%
\bibitem [{\citenamefont {{Hartnoll}}\ \emph {et~al.}(2016)\citenamefont
  {{Hartnoll}}, \citenamefont {{Lucas}},\ and\ \citenamefont
  {{Sachdev}}}]{Hartnoll2016book}%
  \BibitemOpen
  \bibfield  {author} {\bibinfo {author} {\bibfnamefont {S.~A.}\ \bibnamefont
  {{Hartnoll}}}, \bibinfo {author} {\bibfnamefont {A.}~\bibnamefont {{Lucas}}},
  \ and\ \bibinfo {author} {\bibfnamefont {S.}~\bibnamefont {{Sachdev}}},\
  }\href@noop {} {\bibfield  {journal} {\bibinfo  {journal} {arXiv e-prints}\
  ,\ \bibinfo {pages} {arXiv:1612.07324}} (\bibinfo {year} {2016})},\ \Eprint
  {http://arxiv.org/abs/1612.07324} {arXiv:1612.07324 [hep-th]} \BibitemShut
  {NoStop}%
\bibitem [{\citenamefont {Keimer}\ \emph {et~al.}(2015)\citenamefont {Keimer},
  \citenamefont {Kivelson}, \citenamefont {Norman}, \citenamefont {Uchida},\
  and\ \citenamefont {Zaanen}}]{Keimer2015}%
  \BibitemOpen
  \bibfield  {author} {\bibinfo {author} {\bibfnamefont {B.}~\bibnamefont
  {Keimer}}, \bibinfo {author} {\bibfnamefont {S.~A.}\ \bibnamefont
  {Kivelson}}, \bibinfo {author} {\bibfnamefont {M.~R.}\ \bibnamefont
  {Norman}}, \bibinfo {author} {\bibfnamefont {S.}~\bibnamefont {Uchida}}, \
  and\ \bibinfo {author} {\bibfnamefont {J.}~\bibnamefont {Zaanen}},\ }\href
  {\doibase 10.1038/nature14165} {\bibfield  {journal} {\bibinfo  {journal}
  {Nature}\ }\textbf {\bibinfo {volume} {518}},\ \bibinfo {pages} {179}
  (\bibinfo {year} {2015})}\BibitemShut {NoStop}%
\bibitem [{\citenamefont {Maldacena}\ and\ \citenamefont
  {Stanford}(2016)}]{Maldacena2016}%
  \BibitemOpen
  \bibfield  {author} {\bibinfo {author} {\bibfnamefont {J.}~\bibnamefont
  {Maldacena}}\ and\ \bibinfo {author} {\bibfnamefont {D.}~\bibnamefont
  {Stanford}},\ }\href {\doibase 10.1103/PhysRevD.94.106002} {\bibfield
  {journal} {\bibinfo  {journal} {Phys. Rev. D}\ }\textbf {\bibinfo {volume}
  {94}},\ \bibinfo {pages} {106002} (\bibinfo {year} {2016})}\BibitemShut
  {NoStop}%
\bibitem [{\citenamefont {Xu}\ \emph {et~al.}(2017{\natexlab{a}})\citenamefont
  {Xu}, \citenamefont {Sun}, \citenamefont {Schattner}, \citenamefont {Berg},\
  and\ \citenamefont {Meng}}]{xu2017non}%
  \BibitemOpen
  \bibfield  {author} {\bibinfo {author} {\bibfnamefont {X.~Y.}\ \bibnamefont
  {Xu}}, \bibinfo {author} {\bibfnamefont {K.}~\bibnamefont {Sun}}, \bibinfo
  {author} {\bibfnamefont {Y.}~\bibnamefont {Schattner}}, \bibinfo {author}
  {\bibfnamefont {E.}~\bibnamefont {Berg}}, \ and\ \bibinfo {author}
  {\bibfnamefont {Z.~Y.}\ \bibnamefont {Meng}},\ }\href {\doibase
  10.1103/PhysRevX.7.031058} {\bibfield  {journal} {\bibinfo  {journal} {Phys.
  Rev. X}\ }\textbf {\bibinfo {volume} {7}},\ \bibinfo {pages} {031058}
  (\bibinfo {year} {2017}{\natexlab{a}})}\BibitemShut {NoStop}%
\bibitem [{\citenamefont {Berg}\ \emph {et~al.}(2019)\citenamefont {Berg},
  \citenamefont {Lederer}, \citenamefont {Schattner},\ and\ \citenamefont
  {Trebst}}]{Berg2019}%
  \BibitemOpen
  \bibfield  {author} {\bibinfo {author} {\bibfnamefont {E.}~\bibnamefont
  {Berg}}, \bibinfo {author} {\bibfnamefont {S.}~\bibnamefont {Lederer}},
  \bibinfo {author} {\bibfnamefont {Y.}~\bibnamefont {Schattner}}, \ and\
  \bibinfo {author} {\bibfnamefont {S.}~\bibnamefont {Trebst}},\ }\href@noop {}
  {\bibfield  {journal} {\bibinfo  {journal} {Annual Review of Condensed Matter
  Physics}\ }\textbf {\bibinfo {volume} {10}},\ \bibinfo {pages} {63} (\bibinfo
  {year} {2019})}\BibitemShut {NoStop}%
\bibitem [{\citenamefont {Xu}\ \emph {et~al.}(2019{\natexlab{a}})\citenamefont
  {Xu}, \citenamefont {Liu}, \citenamefont {Pan}, \citenamefont {Qi},
  \citenamefont {Sun},\ and\ \citenamefont {Meng}}]{Xu2019}%
  \BibitemOpen
  \bibfield  {author} {\bibinfo {author} {\bibfnamefont {X.~Y.}\ \bibnamefont
  {Xu}}, \bibinfo {author} {\bibfnamefont {Z.~H.}\ \bibnamefont {Liu}},
  \bibinfo {author} {\bibfnamefont {G.}~\bibnamefont {Pan}}, \bibinfo {author}
  {\bibfnamefont {Y.}~\bibnamefont {Qi}}, \bibinfo {author} {\bibfnamefont
  {K.}~\bibnamefont {Sun}}, \ and\ \bibinfo {author} {\bibfnamefont {Z.~Y.}\
  \bibnamefont {Meng}},\ }\href {\doibase 10.1088/1361-648x/ab3295} {\bibfield
  {journal} {\bibinfo  {journal} {Journal of Physics: Condensed Matter}\
  }\textbf {\bibinfo {volume} {31}},\ \bibinfo {pages} {463001} (\bibinfo
  {year} {2019}{\natexlab{a}})}\BibitemShut {NoStop}%
\bibitem [{\citenamefont {Schattner}\ \emph
  {et~al.}(2016{\natexlab{a}})\citenamefont {Schattner}, \citenamefont
  {Lederer}, \citenamefont {Kivelson},\ and\ \citenamefont
  {Berg}}]{Schattner2016}%
  \BibitemOpen
  \bibfield  {author} {\bibinfo {author} {\bibfnamefont {Y.}~\bibnamefont
  {Schattner}}, \bibinfo {author} {\bibfnamefont {S.}~\bibnamefont {Lederer}},
  \bibinfo {author} {\bibfnamefont {S.~A.}\ \bibnamefont {Kivelson}}, \ and\
  \bibinfo {author} {\bibfnamefont {E.}~\bibnamefont {Berg}},\ }\href {\doibase
  10.1103/PhysRevX.6.031028} {\bibfield  {journal} {\bibinfo  {journal} {Phys.
  Rev. X}\ }\textbf {\bibinfo {volume} {6}},\ \bibinfo {pages} {031028}
  (\bibinfo {year} {2016}{\natexlab{a}})}\BibitemShut {NoStop}%
\bibitem [{\citenamefont {Xu}\ \emph {et~al.}(2017{\natexlab{b}})\citenamefont
  {Xu}, \citenamefont {Beach}, \citenamefont {Sun}, \citenamefont {Assaad},\
  and\ \citenamefont {Meng}}]{xu2016topo}%
  \BibitemOpen
  \bibfield  {author} {\bibinfo {author} {\bibfnamefont {X.~Y.}\ \bibnamefont
  {Xu}}, \bibinfo {author} {\bibfnamefont {K.~S.~D.}\ \bibnamefont {Beach}},
  \bibinfo {author} {\bibfnamefont {K.}~\bibnamefont {Sun}}, \bibinfo {author}
  {\bibfnamefont {F.~F.}\ \bibnamefont {Assaad}}, \ and\ \bibinfo {author}
  {\bibfnamefont {Z.~Y.}\ \bibnamefont {Meng}},\ }\href {\doibase
  10.1103/PhysRevB.95.085110} {\bibfield  {journal} {\bibinfo  {journal} {Phys.
  Rev. B}\ }\textbf {\bibinfo {volume} {95}},\ \bibinfo {pages} {085110}
  (\bibinfo {year} {2017}{\natexlab{b}})}\BibitemShut {NoStop}%
\bibitem [{\citenamefont {Liu}\ \emph {et~al.}(2018)\citenamefont {Liu},
  \citenamefont {Xu}, \citenamefont {Qi}, \citenamefont {Sun},\ and\
  \citenamefont {Meng}}]{ZHLiu2018}%
  \BibitemOpen
  \bibfield  {author} {\bibinfo {author} {\bibfnamefont {Z.~H.}\ \bibnamefont
  {Liu}}, \bibinfo {author} {\bibfnamefont {X.~Y.}\ \bibnamefont {Xu}},
  \bibinfo {author} {\bibfnamefont {Y.}~\bibnamefont {Qi}}, \bibinfo {author}
  {\bibfnamefont {K.}~\bibnamefont {Sun}}, \ and\ \bibinfo {author}
  {\bibfnamefont {Z.~Y.}\ \bibnamefont {Meng}},\ }\href {\doibase
  10.1103/PhysRevB.98.045116} {\bibfield  {journal} {\bibinfo  {journal} {Phys.
  Rev. B}\ }\textbf {\bibinfo {volume} {98}},\ \bibinfo {pages} {045116}
  (\bibinfo {year} {2018})}\BibitemShut {NoStop}%
\bibitem [{\citenamefont {Liu}\ \emph {et~al.}(2019{\natexlab{a}})\citenamefont
  {Liu}, \citenamefont {Pan}, \citenamefont {Xu}, \citenamefont {Sun},\ and\
  \citenamefont {Meng}}]{ZHLiu2019}%
  \BibitemOpen
  \bibfield  {author} {\bibinfo {author} {\bibfnamefont {Z.~H.}\ \bibnamefont
  {Liu}}, \bibinfo {author} {\bibfnamefont {G.}~\bibnamefont {Pan}}, \bibinfo
  {author} {\bibfnamefont {X.~Y.}\ \bibnamefont {Xu}}, \bibinfo {author}
  {\bibfnamefont {K.}~\bibnamefont {Sun}}, \ and\ \bibinfo {author}
  {\bibfnamefont {Z.~Y.}\ \bibnamefont {Meng}},\ }\href {\doibase
  10.1073/pnas.1901751116} {\bibfield  {journal} {\bibinfo  {journal}
  {Proceedings of the National Academy of Sciences}\ }\textbf {\bibinfo
  {volume} {116}},\ \bibinfo {pages} {16760} (\bibinfo {year}
  {2019}{\natexlab{a}})}\BibitemShut {NoStop}%
\bibitem [{\citenamefont {Schattner}\ \emph
  {et~al.}(2016{\natexlab{b}})\citenamefont {Schattner}, \citenamefont
  {Gerlach}, \citenamefont {Trebst},\ and\ \citenamefont
  {Berg}}]{Schattner2016a}%
  \BibitemOpen
  \bibfield  {author} {\bibinfo {author} {\bibfnamefont {Y.}~\bibnamefont
  {Schattner}}, \bibinfo {author} {\bibfnamefont {M.~H.}\ \bibnamefont
  {Gerlach}}, \bibinfo {author} {\bibfnamefont {S.}~\bibnamefont {Trebst}}, \
  and\ \bibinfo {author} {\bibfnamefont {E.}~\bibnamefont {Berg}},\ }\href
  {\doibase 10.1103/PhysRevLett.117.097002} {\bibfield  {journal} {\bibinfo
  {journal} {Phys. Rev. Lett.}\ }\textbf {\bibinfo {volume} {117}},\ \bibinfo
  {pages} {097002} (\bibinfo {year} {2016}{\natexlab{b}})}\BibitemShut
  {NoStop}%
\bibitem [{\citenamefont {Gerlach}\ \emph {et~al.}(2017)\citenamefont
  {Gerlach}, \citenamefont {Schattner}, \citenamefont {Berg},\ and\
  \citenamefont {Trebst}}]{Gerlach2017}%
  \BibitemOpen
  \bibfield  {author} {\bibinfo {author} {\bibfnamefont {M.~H.}\ \bibnamefont
  {Gerlach}}, \bibinfo {author} {\bibfnamefont {Y.}~\bibnamefont {Schattner}},
  \bibinfo {author} {\bibfnamefont {E.}~\bibnamefont {Berg}}, \ and\ \bibinfo
  {author} {\bibfnamefont {S.}~\bibnamefont {Trebst}},\ }\href {\doibase
  10.1103/PhysRevB.95.035124} {\bibfield  {journal} {\bibinfo  {journal} {Phys.
  Rev. B}\ }\textbf {\bibinfo {volume} {95}},\ \bibinfo {pages} {035124}
  (\bibinfo {year} {2017})}\BibitemShut {NoStop}%
\bibitem [{\citenamefont {Bauer}\ \emph {et~al.}(2020)\citenamefont {Bauer},
  \citenamefont {Schattner}, \citenamefont {Trebst},\ and\ \citenamefont
  {Berg}}]{Bauer2020}%
  \BibitemOpen
  \bibfield  {author} {\bibinfo {author} {\bibfnamefont {C.}~\bibnamefont
  {Bauer}}, \bibinfo {author} {\bibfnamefont {Y.}~\bibnamefont {Schattner}},
  \bibinfo {author} {\bibfnamefont {S.}~\bibnamefont {Trebst}}, \ and\ \bibinfo
  {author} {\bibfnamefont {E.}~\bibnamefont {Berg}},\ }\href {\doibase
  10.1103/PhysRevResearch.2.023008} {\bibfield  {journal} {\bibinfo  {journal}
  {Phys. Rev. Research}\ }\textbf {\bibinfo {volume} {2}},\ \bibinfo {pages}
  {023008} (\bibinfo {year} {2020})}\BibitemShut {NoStop}%
\bibitem [{\citenamefont {Xu}\ \emph {et~al.}(2019{\natexlab{b}})\citenamefont
  {Xu}, \citenamefont {Qi}, \citenamefont {Zhang}, \citenamefont {Assaad},
  \citenamefont {Xu},\ and\ \citenamefont {Meng}}]{XYXu2019}%
  \BibitemOpen
  \bibfield  {author} {\bibinfo {author} {\bibfnamefont {X.~Y.}\ \bibnamefont
  {Xu}}, \bibinfo {author} {\bibfnamefont {Y.}~\bibnamefont {Qi}}, \bibinfo
  {author} {\bibfnamefont {L.}~\bibnamefont {Zhang}}, \bibinfo {author}
  {\bibfnamefont {F.~F.}\ \bibnamefont {Assaad}}, \bibinfo {author}
  {\bibfnamefont {C.}~\bibnamefont {Xu}}, \ and\ \bibinfo {author}
  {\bibfnamefont {Z.~Y.}\ \bibnamefont {Meng}},\ }\href {\doibase
  10.1103/PhysRevX.9.021022} {\bibfield  {journal} {\bibinfo  {journal} {Phys.
  Rev. X}\ }\textbf {\bibinfo {volume} {9}},\ \bibinfo {pages} {021022}
  (\bibinfo {year} {2019}{\natexlab{b}})}\BibitemShut {NoStop}%
\bibitem [{\citenamefont {Chen}\ \emph {et~al.}(2020)\citenamefont {Chen},
  \citenamefont {Xu}, \citenamefont {Qi},\ and\ \citenamefont
  {Meng}}]{ChuangChen2019}%
  \BibitemOpen
  \bibfield  {author} {\bibinfo {author} {\bibfnamefont {C.}~\bibnamefont
  {Chen}}, \bibinfo {author} {\bibfnamefont {X.~Y.}\ \bibnamefont {Xu}},
  \bibinfo {author} {\bibfnamefont {Y.}~\bibnamefont {Qi}}, \ and\ \bibinfo
  {author} {\bibfnamefont {Z.~Y.}\ \bibnamefont {Meng}},\ }\href {\doibase
  10.1088/0256-307X/37/4/047103} {\bibfield  {journal} {\bibinfo  {journal}
  {Chinese Physics Letters}\ }\textbf {\bibinfo {volume} {37}},\ \bibinfo
  {pages} {047103} (\bibinfo {year} {2020})}\BibitemShut {NoStop}%
\bibitem [{\citenamefont {Assaad}\ and\ \citenamefont
  {Grover}(2016)}]{Assaad2016}%
  \BibitemOpen
  \bibfield  {author} {\bibinfo {author} {\bibfnamefont {F.~F.}\ \bibnamefont
  {Assaad}}\ and\ \bibinfo {author} {\bibfnamefont {T.}~\bibnamefont
  {Grover}},\ }\href {\doibase 10.1103/PhysRevX.6.041049} {\bibfield  {journal}
  {\bibinfo  {journal} {Phys. Rev. X}\ }\textbf {\bibinfo {volume} {6}},\
  \bibinfo {pages} {041049} (\bibinfo {year} {2016})}\BibitemShut {NoStop}%
\bibitem [{\citenamefont {Gazit}\ \emph {et~al.}(2017)\citenamefont {Gazit},
  \citenamefont {Randeria},\ and\ \citenamefont {Vishwanath}}]{Gazit2016}%
  \BibitemOpen
  \bibfield  {author} {\bibinfo {author} {\bibfnamefont {S.}~\bibnamefont
  {Gazit}}, \bibinfo {author} {\bibfnamefont {M.}~\bibnamefont {Randeria}}, \
  and\ \bibinfo {author} {\bibfnamefont {A.}~\bibnamefont {Vishwanath}},\
  }\href {http://dx.doi.org/10.1038/nphys4028} {\bibfield  {journal} {\bibinfo
  {journal} {Nat Phys}\ }\textbf {\bibinfo {volume} {13}},\ \bibinfo {pages}
  {484} (\bibinfo {year} {2017})}\BibitemShut {NoStop}%
\bibitem [{\citenamefont {{Gazit}}\ \emph {et~al.}(2019)\citenamefont
  {{Gazit}}, \citenamefont {{Assaad}},\ and\ \citenamefont
  {{Sachdev}}}]{Gazit2019}%
  \BibitemOpen
  \bibfield  {author} {\bibinfo {author} {\bibfnamefont {S.}~\bibnamefont
  {{Gazit}}}, \bibinfo {author} {\bibfnamefont {F.~F.}\ \bibnamefont
  {{Assaad}}}, \ and\ \bibinfo {author} {\bibfnamefont {S.}~\bibnamefont
  {{Sachdev}}},\ }\href@noop {} {\bibfield  {journal} {\bibinfo  {journal}
  {arXiv e-prints}\ ,\ \bibinfo {eid} {arXiv:1906.11250}} (\bibinfo {year}
  {2019})},\ \Eprint {http://arxiv.org/abs/1906.11250} {arXiv:1906.11250
  [cond-mat.str-el]} \BibitemShut {NoStop}%
\bibitem [{\citenamefont {{Chen}}\ \emph {et~al.}(2020)\citenamefont {{Chen}},
  \citenamefont {{Yuan}}, \citenamefont {{Qi}},\ and\ \citenamefont
  {{Meng}}}]{ChuangChen2020}%
  \BibitemOpen
  \bibfield  {author} {\bibinfo {author} {\bibfnamefont {C.}~\bibnamefont
  {{Chen}}}, \bibinfo {author} {\bibfnamefont {T.}~\bibnamefont {{Yuan}}},
  \bibinfo {author} {\bibfnamefont {Y.}~\bibnamefont {{Qi}}}, \ and\ \bibinfo
  {author} {\bibfnamefont {Z.~Y.}\ \bibnamefont {{Meng}}},\ }\href@noop {}
  {\bibfield  {journal} {\bibinfo  {journal} {arXiv e-prints}\ ,\ \bibinfo
  {eid} {arXiv:2007.05543}} (\bibinfo {year} {2020})},\ \Eprint
  {http://arxiv.org/abs/2007.05543} {arXiv:2007.05543 [cond-mat.str-el]}
  \BibitemShut {NoStop}%
\bibitem [{\citenamefont {Xu}\ \emph {et~al.}(2017{\natexlab{c}})\citenamefont
  {Xu}, \citenamefont {Qi}, \citenamefont {Liu}, \citenamefont {Fu},\ and\
  \citenamefont {Meng}}]{XYXu2017}%
  \BibitemOpen
  \bibfield  {author} {\bibinfo {author} {\bibfnamefont {X.~Y.}\ \bibnamefont
  {Xu}}, \bibinfo {author} {\bibfnamefont {Y.}~\bibnamefont {Qi}}, \bibinfo
  {author} {\bibfnamefont {J.}~\bibnamefont {Liu}}, \bibinfo {author}
  {\bibfnamefont {L.}~\bibnamefont {Fu}}, \ and\ \bibinfo {author}
  {\bibfnamefont {Z.~Y.}\ \bibnamefont {Meng}},\ }\href {\doibase
  10.1103/PhysRevB.96.041119} {\bibfield  {journal} {\bibinfo  {journal} {Phys.
  Rev. B}\ }\textbf {\bibinfo {volume} {96}},\ \bibinfo {pages} {041119}
  (\bibinfo {year} {2017}{\natexlab{c}})}\BibitemShut {NoStop}%
\bibitem [{\citenamefont {Liu}\ \emph {et~al.}(2019{\natexlab{b}})\citenamefont
  {Liu}, \citenamefont {Xu}, \citenamefont {Qi}, \citenamefont {Sun},\ and\
  \citenamefont {Meng}}]{ZHLiu2019EMUS}%
  \BibitemOpen
  \bibfield  {author} {\bibinfo {author} {\bibfnamefont {Z.~H.}\ \bibnamefont
  {Liu}}, \bibinfo {author} {\bibfnamefont {X.~Y.}\ \bibnamefont {Xu}},
  \bibinfo {author} {\bibfnamefont {Y.}~\bibnamefont {Qi}}, \bibinfo {author}
  {\bibfnamefont {K.}~\bibnamefont {Sun}}, \ and\ \bibinfo {author}
  {\bibfnamefont {Z.~Y.}\ \bibnamefont {Meng}},\ }\href {\doibase
  10.1103/PhysRevB.99.085114} {\bibfield  {journal} {\bibinfo  {journal} {Phys.
  Rev. B}\ }\textbf {\bibinfo {volume} {99}},\ \bibinfo {pages} {085114}
  (\bibinfo {year} {2019}{\natexlab{b}})}\BibitemShut {NoStop}%
\bibitem [{\citenamefont {Avraham~Klein}\ and\ \citenamefont
  {Chubukov}(2020)}]{KleinUn2020}%
  \BibitemOpen
  \bibfield  {author} {\bibinfo {author} {\bibfnamefont {E.~B.}\ \bibnamefont
  {Avraham~Klein}, \bibfnamefont {Yoni~Schattner}}\ and\ \bibinfo {author}
  {\bibfnamefont {A.~V.}\ \bibnamefont {Chubukov}},\ }\href
  {https://arxiv.org/abs/2003.09431} {\bibfield  {journal} {\bibinfo  {journal}
  {arXiv:2003.09431 [cond-mat.str-el]}\ } (\bibinfo {year} {2020})}\BibitemShut
  {NoStop}%
\bibitem [{\citenamefont {Bonesteel}\ \emph {et~al.}(1996)\citenamefont
  {Bonesteel}, \citenamefont {McDonald},\ and\ \citenamefont
  {Nayak}}]{Bonesteel1996}%
  \BibitemOpen
  \bibfield  {author} {\bibinfo {author} {\bibfnamefont {N.~E.}\ \bibnamefont
  {Bonesteel}}, \bibinfo {author} {\bibfnamefont {I.~A.}\ \bibnamefont
  {McDonald}}, \ and\ \bibinfo {author} {\bibfnamefont {C.}~\bibnamefont
  {Nayak}},\ }\href {\doibase 10.1103/PhysRevLett.77.3009} {\bibfield
  {journal} {\bibinfo  {journal} {Phys. Rev. Lett.}\ }\textbf {\bibinfo
  {volume} {77}},\ \bibinfo {pages} {3009} (\bibinfo {year}
  {1996})}\BibitemShut {NoStop}%
\bibitem [{\citenamefont {Lederer}\ \emph {et~al.}(2017)\citenamefont
  {Lederer}, \citenamefont {Schattner}, \citenamefont {Berg},\ and\
  \citenamefont {Kivelson}}]{Lederer2017}%
  \BibitemOpen
  \bibfield  {author} {\bibinfo {author} {\bibfnamefont {S.}~\bibnamefont
  {Lederer}}, \bibinfo {author} {\bibfnamefont {Y.}~\bibnamefont {Schattner}},
  \bibinfo {author} {\bibfnamefont {E.}~\bibnamefont {Berg}}, \ and\ \bibinfo
  {author} {\bibfnamefont {S.~A.}\ \bibnamefont {Kivelson}},\ }\href {\doibase
  10.1073/pnas.1620651114} {\bibfield  {journal} {\bibinfo  {journal}
  {Proceedings of the National Academy of Sciences}\ }\textbf {\bibinfo
  {volume} {114}},\ \bibinfo {pages} {4905} (\bibinfo {year}
  {2017})}\BibitemShut {NoStop}%
\bibitem [{\citenamefont {Maslov}\ and\ \citenamefont
  {Chubukov}(2010)}]{Maslov2010}%
  \BibitemOpen
  \bibfield  {author} {\bibinfo {author} {\bibfnamefont {D.~L.}\ \bibnamefont
  {Maslov}}\ and\ \bibinfo {author} {\bibfnamefont {A.~V.}\ \bibnamefont
  {Chubukov}},\ }\href {\doibase 10.1103/PhysRevB.81.045110} {\bibfield
  {journal} {\bibinfo  {journal} {Phys. Rev. B}\ }\textbf {\bibinfo {volume}
  {81}},\ \bibinfo {pages} {045110} (\bibinfo {year} {2010})}\BibitemShut
  {NoStop}%
\bibitem [{\citenamefont {Fradkin}\ \emph {et~al.}(2010)\citenamefont
  {Fradkin}, \citenamefont {Kivelson}, \citenamefont {Lawler}, \citenamefont
  {Eisenstein},\ and\ \citenamefont {Mackenzie}}]{Fradkin2010}%
  \BibitemOpen
  \bibfield  {author} {\bibinfo {author} {\bibfnamefont {E.}~\bibnamefont
  {Fradkin}}, \bibinfo {author} {\bibfnamefont {S.~A.}\ \bibnamefont
  {Kivelson}}, \bibinfo {author} {\bibfnamefont {M.~J.}\ \bibnamefont
  {Lawler}}, \bibinfo {author} {\bibfnamefont {J.~P.}\ \bibnamefont
  {Eisenstein}}, \ and\ \bibinfo {author} {\bibfnamefont {A.~P.}\ \bibnamefont
  {Mackenzie}},\ }\href {\doibase 10.1146/annurev-conmatphys-070909-103925}
  {\bibfield  {journal} {\bibinfo  {journal} {Annual Review of Condensed Matter
  Physics}\ }\textbf {\bibinfo {volume} {1}},\ \bibinfo {pages} {153} (\bibinfo
  {year} {2010})},\ \Eprint
  {http://arxiv.org/abs/https://doi.org/10.1146/annurev-conmatphys-070909-103925}
  {https://doi.org/10.1146/annurev-conmatphys-070909-103925} \BibitemShut
  {NoStop}%
\bibitem [{\citenamefont {Raghu}\ \emph {et~al.}(2015)\citenamefont {Raghu},
  \citenamefont {Torroba},\ and\ \citenamefont {Wang}}]{Raghu2015}%
  \BibitemOpen
  \bibfield  {author} {\bibinfo {author} {\bibfnamefont {S.}~\bibnamefont
  {Raghu}}, \bibinfo {author} {\bibfnamefont {G.}~\bibnamefont {Torroba}}, \
  and\ \bibinfo {author} {\bibfnamefont {H.}~\bibnamefont {Wang}},\ }\href
  {\doibase 10.1103/PhysRevB.92.205104} {\bibfield  {journal} {\bibinfo
  {journal} {Phys. Rev. B}\ }\textbf {\bibinfo {volume} {92}},\ \bibinfo
  {pages} {205104} (\bibinfo {year} {2015})}\BibitemShut {NoStop}%
\bibitem [{\citenamefont {Metlitski}\ \emph {et~al.}(2015)\citenamefont
  {Metlitski}, \citenamefont {Mross}, \citenamefont {Sachdev},\ and\
  \citenamefont {Senthil}}]{Metlitski2015}%
  \BibitemOpen
  \bibfield  {author} {\bibinfo {author} {\bibfnamefont {M.~A.}\ \bibnamefont
  {Metlitski}}, \bibinfo {author} {\bibfnamefont {D.~F.}\ \bibnamefont
  {Mross}}, \bibinfo {author} {\bibfnamefont {S.}~\bibnamefont {Sachdev}}, \
  and\ \bibinfo {author} {\bibfnamefont {T.}~\bibnamefont {Senthil}},\ }\href
  {\doibase 10.1103/PhysRevB.91.115111} {\bibfield  {journal} {\bibinfo
  {journal} {Phys. Rev. B}\ }\textbf {\bibinfo {volume} {91}},\ \bibinfo
  {pages} {115111} (\bibinfo {year} {2015})}\BibitemShut {NoStop}%
\bibitem [{\citenamefont {Lederer}\ \emph {et~al.}(2015)\citenamefont
  {Lederer}, \citenamefont {Schattner}, \citenamefont {Berg},\ and\
  \citenamefont {Kivelson}}]{Lederer2015}%
  \BibitemOpen
  \bibfield  {author} {\bibinfo {author} {\bibfnamefont {S.}~\bibnamefont
  {Lederer}}, \bibinfo {author} {\bibfnamefont {Y.}~\bibnamefont {Schattner}},
  \bibinfo {author} {\bibfnamefont {E.}~\bibnamefont {Berg}}, \ and\ \bibinfo
  {author} {\bibfnamefont {S.~A.}\ \bibnamefont {Kivelson}},\ }\href {\doibase
  10.1103/PhysRevLett.114.097001} {\bibfield  {journal} {\bibinfo  {journal}
  {Phys. Rev. Lett.}\ }\textbf {\bibinfo {volume} {114}},\ \bibinfo {pages}
  {097001} (\bibinfo {year} {2015})}\BibitemShut {NoStop}%
\bibitem [{\citenamefont {Abrikosov}\ \emph {et~al.}(1975)\citenamefont
  {Abrikosov}, \citenamefont {Gorkov},\ and\ \citenamefont
  {Dzyaloshinski}}]{Abrikosov1975}%
  \BibitemOpen
  \bibfield  {author} {\bibinfo {author} {\bibfnamefont {A.}~\bibnamefont
  {Abrikosov}}, \bibinfo {author} {\bibfnamefont {L.}~\bibnamefont {Gorkov}}, \
  and\ \bibinfo {author} {\bibfnamefont {I.}~\bibnamefont {Dzyaloshinski}},\
  }\href@noop {} {\emph {\bibinfo {title} {Methods of Quantum Field Theory in
  Statistical Physics}}},\ Dover Books on Physics Series\ (\bibinfo
  {publisher} {Dover Publications},\ \bibinfo {year} {1975})\BibitemShut
  {NoStop}%
\bibitem [{Note1()}]{Note1}%
  \BibitemOpen
  \bibinfo {note} {As shown in Ref.~\cite {xu2017non}, it turns out the bosonic
  propagator is dominated by the bosonic self-energy part, with a small finite
  anomalous dimension in $q^2$ and $\Omega ^2$ terms, it will not change the
  main results of this paper.}\BibitemShut {Stop}%
\bibitem [{\citenamefont {Chubukov}\ and\ \citenamefont
  {Maslov}(2012)}]{Chubukov2012}%
  \BibitemOpen
  \bibfield  {author} {\bibinfo {author} {\bibfnamefont {A.~V.}\ \bibnamefont
  {Chubukov}}\ and\ \bibinfo {author} {\bibfnamefont {D.~L.}\ \bibnamefont
  {Maslov}},\ }\href {\doibase 10.1103/PhysRevB.86.155136} {\bibfield
  {journal} {\bibinfo  {journal} {Phys. Rev. B}\ }\textbf {\bibinfo {volume}
  {86}},\ \bibinfo {pages} {155136} (\bibinfo {year} {2012})}\BibitemShut
  {NoStop}%
\bibitem [{\citenamefont {Chubukov}(2005{\natexlab{b}})}]{Chubukov2005}%
  \BibitemOpen
  \bibfield  {author} {\bibinfo {author} {\bibfnamefont {A.~V.}\ \bibnamefont
  {Chubukov}},\ }\href {\doibase 10.1103/PhysRevB.72.085113} {\bibfield
  {journal} {\bibinfo  {journal} {Phys. Rev. B}\ }\textbf {\bibinfo {volume}
  {72}},\ \bibinfo {pages} {085113} (\bibinfo {year}
  {2005}{\natexlab{b}})}\BibitemShut {NoStop}%
\bibitem [{\citenamefont {Schlief}\ \emph {et~al.}(2017)\citenamefont
  {Schlief}, \citenamefont {Lunts},\ and\ \citenamefont {Lee}}]{SKLee2017}%
  \BibitemOpen
  \bibfield  {author} {\bibinfo {author} {\bibfnamefont {A.}~\bibnamefont
  {Schlief}}, \bibinfo {author} {\bibfnamefont {P.}~\bibnamefont {Lunts}}, \
  and\ \bibinfo {author} {\bibfnamefont {S.-S.}\ \bibnamefont {Lee}},\ }\href
  {\doibase 10.1103/PhysRevX.7.021010} {\bibfield  {journal} {\bibinfo
  {journal} {Phys. Rev. X}\ }\textbf {\bibinfo {volume} {7}},\ \bibinfo {pages}
  {021010} (\bibinfo {year} {2017})}\BibitemShut {NoStop}%
\end{thebibliography}%
\end{document}